\documentclass[prd,aps,amssymb,amsmath,tightenlines,nofootinbib]{revtex4}
\usepackage{graphicx,epsfig}
\usepackage{slashed}
\usepackage[colorinlistoftodos]{todonotes}
\usepackage{appendix}

\newcommand{\be}{\begin{equation}}
\newcommand{\ee}{\end{equation}}

\begin{document}

\preprint[{\leftline{KCL-PH-TH/2022-{\bf 58}}

\title{Axion effective potentials induced by heavy sterile fermions 
}

\author{Nick E. Mavromatos$^{a,b}$}

\author{Sarben Sarkar$^b$ }

\affiliation{$^a$Physics Department, School of Applied Mathematical and Physical Sciences, National Technical University of Athens , Athens 157 80, Greece}

\affiliation{$^b$Theoretical Particle Physics and Cosmology Group,\\ Department of Physics, King's College London,\\
London, WC2R 2LS, UK}

\begin{abstract}
A model of (3+1)-dimensional leptogenesis, proposed previously by the authors, requires a CPT Violating (CPTV) background of the Kalb-Ramond (KR) axion field. {The KR axion is} a pseudoscalar, which is dual to the field strength of the spin-one field present in the massless gravitational multiplet in the theory of closed bosonic strings (compactified to four dimensions). Microscopic models for the emergence of such backgrounds are provided {both} by low-energy {string-inspired} gravitational theories with torsion (including (primordial) gravitational and axial gauge anomalies) and  {by Einstein-Cartan gravity, a closely related simpler model.}
In this work we examine the pseudoscalar  \emph{quanta} of the KR axion in this background using the methods of effective field theory. {In our model for leptogenesis there is}  
a single sterile {right-handed} neutrino (RHN) with mass $m_N$. At energies lower than $m_N$, an axion potential is derived by integrating out {at one loop} the sterile neutrino in the spirit of effective field theory. The stability of this axion potential is important for the viability of our model. The vacuum of this potential is generally metastable. The stability of the vacuum is determined by the ratio 
of the torsion-induced-axion coupling (which depends on the string mass scale) to $m_N$, which should be larger or equal to one, for the validity of our effective field theory. The vacuum is metastable only for axion couplings much larger than the sterile neutrino mass (large string mass scales, e.g. comparable to the four-dimensional Planck mass), with a lifetime much larger than the age of the observable Universe.  By contrast, if axion couplings are comparable to the RHN mass the false vacuum is highly unstable, with a lifetime much smaller than the age of the observable Universe; in this case the CPTV leptogenesis scenario is not viable.

\end{abstract}


\maketitle
\section{Introduction: 
Unconventional CPT-Violating Leptogenesis }\label{aba:sec1}
The observed baryon asymmetry (i.e. the fact that there is hardly any primordial antimatter) in the current cosmological era of  the Universe~\cite{Planck} implies that, at an earlier era (for times $t<10^{-6}$s),
\begin{equation}
\label{e1}
\Delta n_{B}\left(T\sim ~{\rm 1~GeV}\right)=\frac{n_{B}-n_{\overline{B}}}{n_{B}+n_{\overline{B}}}\sim(8.4-8.9)\times10^{-11}
\end{equation}
where $n_B$ is the baryon number density and $n_{\overline{B}}$ is the anti-baryon number density at {at this earlier epoch} 
of the universe. An asymmetry $\Delta n_{L}$ between leptons and anti-leptons\footnote{Lepton asymmetry is hard to measure directly.} is also expected to be of the same order of magnitude as $\Delta n_{B}$ (i.e. there are hardly any antineutrinos compared to neutrinos). The chiral anomaly in the Standard Model (SM) is absent for $\Delta n_{B}-\Delta n_{L} = 0$, which correlates leptogenesis and baryogengesis.

 In {SM} it is \emph{not} possible to find the required amount of baryon asymmetry in \eqref{e1}.  For CPT~\cite{Streater:1989vi,schwartz,R1b} invariant theories, Sakharov~\cite{Sakharov:1967dj}  states  the following criteria   for baryogenesis :
\begin{enumerate}
  \item The existence of  baryon-number violating processes in the SM (which happen to be non-perturbative)
  \item C and CP violation~\cite{Bigi:2000yz} should be present so that the amplitudes for the processes 
  $X\rightarrow Y+b$ and $\bar{X}\rightarrow\bar{Y}+\bar{b}$ are unequal ( where $X, Y$ and $b$ denote particles and the barred quantities their antiparticles).
   \item  Existence of processes out of chemical equilibrium
\end{enumerate}
The observed CP violation in the neutral kaon experiments is insufficient to obtain the result of (\ref{e1}). There is no consensus on the way to increase the amount of CP violation~\cite{Shaposhnikov:2009zzc},  since they involve  fine tuning especially in the parameters entering the one-loop diagrams associated with leptogenesis/baryogenesis; so the framework of Sakharov remains qualitative.

Baryogenesis and leptogenesis require explanations which lie beyond SM (BSM). The existence of neutrino masses also requires BSM.There is an approach to leptogenesis which builds on the see-saw model for neutrino masses~~\cite{Fukugita,Mohapatra,Davidson:2008bu,Covi:1996wh,Buchmuller:2004nz,Buchmuller:2005eh,Pilaftsis:2003gt,Pilaftsis:2005rv} and so connects two pieces of BSM physics. {Gravity is also BSM physics} and is addressed by String Theory, a framework for BSM physics. We propose a solution to baryogengesis  through a solution to leptogenesis incorporating gravitational physics derived from String theory. Sphaleron processes~\cite{Gorbunov} then convert lepton number asymmetry into baryogenesis~\cite{Luty}. In such a scenario lepton asymmetry is generated first by means of decays of right handed sterile Majorana neutrinos to SM particles which freeze out at a certain temperature because of the expanding universe. 
In the \emph{conventional} leptogenesis scenarios, within the framework of flat spacetime and CPT invariance, the effective field theories require fine tuning of the parameters affecting the loop diagrams of the heavy neutrino decays (in the Yukawa sector). 

In an attempt to avoid this fine tuning problem, we discuss in \cite{Mavromatos:2012ii,deCesare:2014dga,Ellis:2013gca,R10Boss,R11Boss,R12Mav,R13Mav,MavRS} scenarios where a matter-antimatter asymmetry in late eras is due to an appropriate non-Minkowski geometry in the early Universe. This geometry is characterised by spontaneous Lorentz (LV) and (associated) CPT Violation (CPTV), due to the existence of appropriate background fields of axion fields. In string theory these backgrounds cannot be arbitrary because of the requirements of conformal invariance.  The axion fields are dual to torsion in the effective four-dimensional gravitational theories obtained from strings~\cite{Green:2012oqa,Green:2012pqa,Duncan:1992vz,Polchinski:1998rq,Polchinski:1998rr}.\footnote{We are not able to justify the restriction of degrees of freedom to the gravitational multiplet found in closed bosonic string theory because string theory is a framework which entails many degrees of freedom. Hence our approach is a phenomenological model inspired by string theory.} 
Such scenarios appear viable from a phenomenological point of view and can be mapped to the Standard Model Extension (SME)~\cite{Colladay:1998fq,Kostelecky:2008ts} (see also \cite{Mavromatos:2013gya}) which provides a framework for studies of the breaking of Lorentz and CPT symmetries.

In this work we wish to extend the reach of our CPTV Leptogenesis model to encompass axionic dark matter, and quantum torsion degrees of freedom which could be linked to axion-like particles. On integrating out heavy right-handed sterile neutrinos (RHN), one obtains a shift-symmetry-breaking effective potential of the initially massless gravitational axions (existing in the string spectrum). Applying effective field theory methods, we establish conditions under which the (one-loop) effective axion potential leads to a non-trivial mass term, thereby making the gravitational axion  a potential dark matter candidate. However, in order to establish that such axions are viable as dark matter candidates, we need to check the stability of the axion vacuum, which is determined through the corresponding one-loop effective potential. We find that for high string mass scales, compared to the RHN mass, the vacuum is metastable, with a life time much larger than the age of the Universe. This establishes that the gravitational axion of the CPTV Leptogenesis model gives a viable candidate for dark matter. Given the geometric interpretation of the gravitational axion as torsion in string inspired effective gravitational field theories, we thus establish a geometrical origin of dark matter.

The structure of this article is the following: in section \ref{sec:cptvlept} we review for completion the main features of the CPTV leptogenesis model. In section \ref{sec:cptvleptscen}, we review the relationship between geometry and axion-like particles in both, string effective theories and Einstein-Cartan framework, which generically contains torsional spacetime. We also perform a brief, comparative phenomenological study of these two frameworks. In section \ref{sec:stable} we construct the effective gravitational-axion potential, obtained from integrating out heavy RHN degrees of freedom, and show that for high string mass scales the resulting vacuum is metastable, with a life time much larger than the age of the Universe. Finally, conclusions and outlook are given in section \ref{sec:concl}. Some technical aspects regarding estimates of the vacuum life time are given in an Appendix.

\section{The leptogenesis model}\label{sec:cptvlept}

 In \cite{deCesare:2014dga, R10Boss} we  consider leptogenesis originating from the \emph{tree-level} decay of a heavy sterile (right-handed, Majorana) neutrino (RHN)  into SM leptons, in the presence of generic \emph{CPTV time-like axial backgrounds}, assumed constant in the cosmological (Robertson-Walker) frame of the early universe\footnote{For sigma models in bosonic string theories such backgrounds are known to be solutions consistent with conformnal invariance required in the formulation of string theory~\cite{Antoniadis:1988vi}. }. The resulting Lagrangian is given by: 
\begin{equation}
\label{e2}
{\mathcal{L}}= {\mathcal L}_{\rm SM} + i\overline{N}\slashed{\partial}N-\frac{m_{N}}{2}(\overline{N^{c}}N+\overline{N}N^{c})-\overline{N}\slashed{B}\gamma^{5}N-\sum_k \, y_{k}\overline{L}_{k}\widetilde{\varphi}N+h.c.
\end{equation}
where ${\mathcal L}_{\rm SM}$ denotes the SM Lagrangian, $B_\mu$ is a CPTV background c-number {BSM} field, associated with physics beyond the SM, 
$N$ is the RHN field with (Majorana) mass $m_N$,  $\widetilde \varphi$ is the adjoint ($\widetilde{\varphi}_i=\varepsilon_{ij}\varphi_j $) of the Higgs field  $\varphi$, 
 and $L_{k}$ is a lepton (doublet) field of the SM sector, with $k$ a generation index. $y_k$ is a Yukawa coupling, which is non-zero and provides a non-trivial (``Higgs portal'') interaction between the RHN and the SM sectors. In ~\cite{deCesare:2014dga, R10Boss} a single sterile neutrino species is shown
to generate phenomenologically relevant lepton asymmetry. From now on 
we restrict ourselves to $k=1$, and set 
\begin{equation}
\label{e3}
y_1 = y ~.
\end{equation}
In the scenario of \cite{R9, R10Boss}, the CPTV background $B_\mu$ is assumed to have only a non-zero temporal component~\cite{Antoniadis:1988vi}, which is taken to be constant in the Robertson-Walker frame of the early universe, 
\begin{equation}\label{temporalB}
B_0 = {\rm const} \ne 0~, \, B_i = 0 ~, i=1,2,3~.
\end{equation}
In this case, the Lagrangian (\ref{e2}) assumes the form of a SME Lagrangian in a LV and CPTV background~\cite{Colladay:1998fq,Kostelecky:2008ts,Mavromatos:2013gya}. 

A lepton asymmetry is then generated due to the CP and CPTV tree-level decays of  $N$ into SM leptons, 
in the presence of the background (\ref{temporalB}), induced by the Higgs portal Yukawa interactions of (\ref{e2})~\cite{Colladay:1998fq,R10Boss}:
\begin{eqnarray}\label{4channels}
{\rm Channel ~I}&:& \qquad  N \rightarrow l^{-}h^{+}~, ~ \nu \, h^{0}~,  \\ \nonumber 
{\rm Channel ~II}&:& \qquad  N \rightarrow l^{+}h^{-}~,~  \overline \nu \, h^{0}~.
\end{eqnarray}
where $\ell^\pm$ are charged leptons, $\nu$ ($\overline \nu$) are light, ``active'', neutrinos (antineutrinos) in the SM sector,
$h^0$ is the neutral Higgs field, and 
 $h^\pm$ are the charged Higgs fields\footnote{At high temperatures, above the spontaneous electroweak symmetry breaking, the charged Higgs fields $h^\pm$ do not decouple from the physical spectrum, and play an important r\^ole in leptogenesis.}.  As a result of the non-trivial $B_0 \ne 0$ background (\ref{temporalB}), the decay rates of the Majorana RHN between the channels I and II are different, resulting in a Lepton asymmetry, $\Delta L^{TOT}$, which then freezes out a temperature $T_D$. In \cite{R10Boss, R11Boss}, a detailed study of the associated Boltzmann equations for the processes in (\ref{4channels}), and their reciprocals,  has led to the result:
\begin{equation}\label{totDL}
\dfrac{\Delta L^{TOT}}{s} \simeq  (0.016, \, 0.019) \,  \dfrac{B_{0}}{m_{N}},  
\end{equation}
$ {\rm at~the~ freezeout~temperature} \quad T=T_D : \quad m_N/T_D  \simeq (1.44, \, 1.77),$where $s$ is the entropy density of the universe (and the numbers inside the  brackets distinguish different Pad$\acute{e}$ approximations, see discussion below). This implies the phenomenologically acceptable values of the lepton asymmetry of ${\mathcal O}(8 \times 10^{-11})$, which can then be communicated to the baryon sector via Baryon-minus-Lepton-number ($B-L$) conserving sphaleron processes~\cite{R13Mav} in the SM, thus producing the observed amount of baryogenesis  in the Universe,  for values of
\begin{equation}\label{b0}
\frac{B_0}{m_N} \sim  10^{-9}, \quad {\rm at~ freezeout~temperature} \quad T=T_D : \quad m_N/T_D  \simeq (1.77, 1.44),
\end{equation}
The different values $(a,b)$ of the numerical coefficients in the right-hand-side of (\ref{totDL}), are due to two different analytical methods (series expansion ($a$) and integrating factor ($b$) method~\cite{R10Boss, R11Boss}, respectively) used in the Pad$\acute{e}$ approximant solution of the Boltzmann equations associated with (\ref{4channels}). With the value $y \sim 10^{-5}$ of the Yukawa coupling (\ref{e3}) , and for $m_N = {\mathcal O}(100)$~TeV~\cite{deCesare:2014dga, R10Boss} we  obtain a $B_0 \sim 0.1~{\rm MeV}$, for phenomenologically relevant leptogenesis to occur at $T_D \simeq (56 - 69) $ TeV, in our scenario.
In \cite{deCesare:2014dga, R10Boss} the microscopic origin of the background $B_0$ was not discussed in detail.

\section{Anomalies, Torsion and microscopic scenarios for CPTV Leptogenesis}\label{sec:cptvleptscen}

The Lagrangian \eqref{e2} is a phenomenological one, in which the microscopic origin of the $B_\mu$ axial background is not specified. There is a microscopic origin of $B_\mu$. It can arise in two ways:
\begin{itemize}
    \item as \emph{ string-inspired effective  gravitational theories}~(SEG)~\cite{deCesare:2014dga,bms1,bms2,ms1,ms2,Sarkar:2022odh,mavtorsion}, characterised by chiral gravitational anomalies and torsion in the early Universe~\cite{Green:2012oqa,Duncan:1992vz,Polchinski:1998rq,Polchinski:1998rr},
    \item as generic 
Einstein-Cartan theories~\cite{Cartan2001RiemannianGI,Utiyama:1956sy,Kibble:1961ba} with condensates of (the totally antisymmetric part of the) torsion, which we describe briefly in this section. 
\end{itemize}
The non-conservation of the axial current for the RHN due to the chiral anomaly~\cite{Adler:1969gk,Fujikawa:1979ay} is crucial for  CPTV leptogenesis.
We commence our discussion with the more fundamental string-inspired case in the next subsection. Subsequently we show that the Einstein-Cartan model has closely related features but has much simplified formulation. We briefly compare and contrast the phenomenology of the two models.

\subsection{String-inspired Chern-Simons Gravitational models}\label{sec:CSmodels}

Effective gravitational theories arise from strings in the limit of low-energies compared to the string energy scale; torsion consists of a totally antisymmetric component only, which is provided by the field strength $\mathcal H_{\mu\nu\rho}$~\cite{Gross:1986mw,Metsaev:1987zx,Bento:1986hx,Green:2012oqa} of the Kalb-Ramond (KR) antisymmetric tensor field~\cite{Kalb}~\footnote{Together with the dilaton and graviton, this field, of spin one, constitutes the massless bosonic string gravitational multiplet.}. The field strength  is~ modified by the Green-Schwarz terms~\cite{gs}, required for cancellation of  the extra-dimensional gauge and gravitational anomalies in the underlying string theory.  This modification requires the addition of  Chern-Simons (CS)  (gravitational (``Lorentz'', L)  and gauge (Y)) anomalous terms  to the curl of the KB field  :
\begin{align}\label{GSH}
\mathbf{H} = \mathbf{d} \mathbf{B} \quad \Rightarrow \quad \mathbf{{\mathcal {\bf{H}}}} &= \mathbf{d} \mathbf{B} + \frac{\alpha^\prime}{8\, \kappa} \, \Big({\boldsymbol{\Omega}}_{\rm 3L} - {\boldsymbol{\Omega}}_{\rm 3Y}\Big),  \nonumber \\
{\boldsymbol{{\boldsymbol{\Omega}}}}_{\rm 3L} &= {\boldsymbol{\omega}}^a_{\,\,c} \wedge \mathbf{d} {\boldsymbol{\Omega}}^c_{\,\,a}
+ \frac{2}{3}  {\boldsymbol{\omega}}^a_{\,\,c} \wedge  {\boldsymbol{\omega}}^c_{\,\,d} \wedge {\boldsymbol{\Omega}}^d_{\,\,a},
\quad {\boldsymbol{\Omega}}_{\rm 3Y} = \mathbf{A} \wedge  \mathbf{d} \mathbf{A} + \mathbf{A} \wedge \mathbf{A} \wedge \mathbf{A},
\end{align}
where we use the language of differential forms~\cite{alma99900524832201854} for notational brevity.

In \eqref{GSH} $\mathcal H$ is a three-form, 
the symbol $\wedge$ denotes the exterior product of differential ($k,\ell$) forms (${\mathbf f}^{(k)} \wedge {\mathbf g}^{(\ell)} = (-1)^{k\, \ell}\, {\mathbf g}^{(\ell)} \wedge {\mathbf f}^{(k)}$), $\mathbf{A} \equiv \mathbf A_\mu \, dx^\mu$ denotes the one form for the Yang-Mills gauge field, and $\omega^a_{\,\,b} \equiv \omega^a_{\,\,\,\mu\,b}\, dx^\mu$ is the one form for the spin connection, where the Latin indices $a,b,c,d$ are  ($SO(1,3$)) indices of the tangent space. The quantity $\alpha^\prime$ is the Regge slope $\alpha^\prime=M_s^{-2}$, where 
$M_s$ is the string mass scale. $M_s$ differs from the reduced four dimensional Planck scale  $M_{\rm Pl}(=  2.43 \times 10^{18}$~GeV   in natural units $\hbar=c=1$)
. The four-dimensional gravitational
constant $\kappa = \sqrt{8\pi\, {\rm G}} = M_{\rm Pl}^{-1}$. 

The modification \eqref{GSH} implies the Bianchi identity
\begin{align}\label{modbianchi2}
 \varepsilon_{\alpha\beta\sigma}^{\;\;\;\;\;\;\;\;\mu}\, {\mathcal H}^{\alpha\beta\sigma}_{\;\;\;\;\;\;\; ;\mu}
 =  \frac{\alpha^\prime}{32\, \kappa} \, \sqrt{-g}\, \Big(R_{\mu\nu\rho\sigma}\, \widetilde R^{\mu\nu\rho\sigma} - \mathbf F_{\mu\nu} \, \widetilde{\mathbf F}^{\mu\nu}\Big)\,,
 \end{align}
where the semicolon \textquotedblleft $; $\textquotedblright denotes the gravitational covariant derivative with respect to the standard (torsion-free)
Christoffel connection,  $\mathbf{F} _{\mu\nu}$ is the non-Abelian gauge field strength, $R_{\mu\nu\rho\sigma}$ is the four-dimensional Riemann tensor, and $ \widetilde R^{\mu\nu\rho\sigma} = \frac{1}{2} \varepsilon^{\mu\nu\alpha\beta} \, R_{\alpha\beta}^{\quad\, \rho\sigma}$, $ \widetilde{\mathbf F}^{\mu\nu} = \frac{1}{2} \varepsilon^{\mu\nu\alpha\beta} \, \mathbf{F}_{\alpha\beta}$ are the corresponding dual tensors, 
with 
\begin{align}\label{coveps}
\varepsilon_{\mu\nu\rho\sigma} = \sqrt{-g}\,  \epsilon_{\mu\nu\rho\sigma}\,, \quad \, \varepsilon^{\mu\nu\rho\sigma} =\frac{{\rm sgn}(g)}{\sqrt{-g}}\,  \epsilon^{\mu\nu\rho\sigma}\,, 
\end{align}
with $g$ the metric determinant, and  $\epsilon_{\mu\nu\rho\sigma}$ ($\epsilon_{0123} = +1$, {\emph etc.}) is the Minkowski-space-time (totally antisymmetric) Levi-Civita  symbol.\footnote{In this work we follow the convention
for the signature of the metric $(+, -,-,- )$, and the definitions of the Riemann Curvature tensor
$R^\lambda_{\,\,\,\,\mu \nu \sigma} = \partial_\nu \, \Gamma^\lambda_{\,\,\mu\sigma} + \Gamma^\rho_{\,\, \mu\sigma} \, \Gamma^\lambda_{\,\, \rho\nu} - (\nu \leftrightarrow \sigma)$, the Ricci tensor $R_{\mu\nu} = R^\lambda_{\,\,\,\,\mu \lambda \nu}$, and the Ricci scalar $R = R_{\mu\nu}g^{\mu\nu}$.} 
The right-hand side of \eqref{modbianchi2} corresponds to the four-dimensional mixed CS (gravitational and gauge) anomaly~\cite{Jackiw:2003pm}, and is a total derivative of the topological anomalous current density $\mathcal K^\mu$:
\begin{align}\label{pontryaginA}
&\sqrt{-g} \, \Big(R_{\mu\nu\rho\sigma}\, \widetilde R^{\mu\nu\rho\sigma} - \mathbf{F}_{\mu\nu}\,\widetilde{\mathbf{F}}^{\mu\nu} \Big) = \sqrt{-g} \, {\mathcal K}^\mu (\omega)_{;\mu} = \partial_\mu \Big(\sqrt{-g} \, {\mathcal K}^\mu (\omega) \Big) \nonumber \\
&= 2 \, \partial_\mu \Big[\epsilon^{\mu\nu\alpha\beta}\, \omega_\nu^{ab}\, \Big(\partial_\alpha \, \omega_{\beta ab} + \frac{2}{3}\, \omega_{\alpha a}^{\,\,\,\,\,\,\,c}\, \omega_{\beta cb}\Big) - 2 \epsilon^{\mu\nu\alpha\beta}\, \Big(A^i_\nu\, \partial_\alpha A_\beta^i + \frac{2}{3} \, f^{ijk} \, A_\nu^i\, A_\alpha^j \, A_\beta^k \Big)\Big],
\end{align}
with Latin letters $i,j,k$ being gauge group indices.

The torsion interpretation of the field strength $\mathcal H$~\cite{Gross:1986mw,Metsaev:1987zx,Bento:1986hx,Green:2012oqa}, implies a \emph{linear coupling} of $\mathcal H$ with the \emph{axial fermion} current $\sum_i \overline \psi_i \gamma^\mu \, \gamma^5 \, \psi$, where the sum is over {\it all } fermion species of the theory: the kinetic terms of the fermions contain the contorted gravitational covariant derivative which is \emph{linear} in the contorsion~\cite{Cartan2001RiemannianGI,torsion,Shapiro:2001rz}. The Bianchi identity,  \eqref{modbianchi2} is implemented~ \cite{Duncan:1992vz,svrcek}, in the effective action \eqref{sea6},  by means of a (canonically normalised) pseudoscalar Lagrange multiplier field $b(x)$; the path-integration over the torsion $H$-field yields~\cite{Duncan:1992vz,svrcek,deCesare:2014dga,Mavromatos:2012ii,bms1}:  
 \begin{align}\label{sea6}
&S^{\rm eff} =\; \int d^{4}x\sqrt{-g}\Big[ -\dfrac{1}{2\kappa^{2}}\, R + \frac{1}{2}\, \partial_\mu b \, \partial^\mu b -  \sqrt{\frac{2}{3}}\,
\frac{\alpha^\prime}{96\, \kappa} \, \partial_\mu b(x) \, {\mathcal K}^\mu
\Big]  \nonumber \\
&+ S_{\rm Dirac~or~Majorana}^{Free} + \int d^{4}x\sqrt{-g}\, \Big[\Big( {\mathcal F}_\mu + \frac{\alpha^\prime}{2\, \kappa} \, \sqrt{\frac{3}{2}} \, \partial_{\mu}b \Big)\, J^{5\mu}    - \dfrac{3\alpha^{\prime\, 2}}{16 \, \kappa^2}\,J^{5}_{\mu}J^{5\mu} \Big] + \dots\,,
\end{align}
where we use \eqref{pontryaginA}; $J^{5 \mu}~\equiv~\sum_i \, \overline \psi \gamma^5 \, \gamma^\mu \, \psi $ denotes 
the axial fermion current; ${\mathcal F}^d  =   \varepsilon^{abcd} \, e_{b\lambda} \,  \partial_a \, e^\lambda_{\,\,c}$ with $e^\mu_{\,\,c}$ the vielbeins~\footnote{Latin indices pertain to the tangent space of the space-time manifold at a given point.};  
$S_{\rm Dirac~or~Majorana}^{Free}$ denotes the free-fermion kinetic terms; 
 the $\dots$ in (\ref{sea6}) indicate gauge field kinetic terms, as well as terms of higher order in derivatives. The action \eqref{sea6} is valid for both Dirac or Majorana fermions. The model described by \eqref{sea6} is a \emph{CS gravity} model~\cite{Jackiw:2003pm} with massless axions $b(x)$ and fermions.

 We note the presence in \eqref{sea6} of the CP-violating interactions of the derivative of the field $b$ with the axial fermion current $J^{5\mu} $. The coupling in this interaction defines the  axion coupling, which in this case turns out to be:\footnote{In string theory, there are also other axion fields arising from compactification, which depend on the specific string model considered~\cite{svrcek}. Such multiaxion situations can lead to a very interesting and rich string phenomenology~\cite{axiverse,Marsh:2015xka,Mehta:2021pwf,Mehta:2020kwu}. }
\begin{align}\label{kraxcoupl}
f_b^{\rm str} = 96\, \sqrt{\frac{3}{2}} \, \frac{\kappa}{\alpha^\prime} \simeq 2 \times 10^2  \, \frac{\kappa}{\alpha^\prime}\,.
\end{align}
Moreover, integration over the quadratic $H$-torsion terms, yields the  
{\it repulsive} four-fermion axial-current-current interaction term, 
\begin{align}\label{4fint}
\mathcal S_{\rm 4f-int} = -\int d^4x \, \sqrt{-g}\, \dfrac{3\alpha^{\prime\, 2}}{16 \, \kappa^2}\,J^{5}_{\mu}J^{5\mu}\,, 
\end{align}
which is also characteristic of theories with spin-1/2 Einstein-Cartan torsion~\cite{Cartan2001RiemannianGI,Shapiro:2001rz,Sarkar:2022odh}. In fact the Lagrangian \eqref{sea6}, \emph{except} for the primordial anomaly term involving $\mathcal K^\mu$, coincides with the generic Einstein-Cartan Lagrangian for fermions upon restricting ourselves to the case of \emph{totally antisymmetric torsion}~\cite{Duncan:1992vz}, which we shall review briefly in \ref{EC}, for completeness.

\subsubsection{The stringy running vacuum model}

 The term  proportional to $\mathcal K^\mu$ in~\eqref{sea6} is absent in the Einstein-Cartan model and will not play a direct role in our analysis. It is indirectly useful however, in the stringy running vacuum model (RVM)\cite{bms1,bms2,ms1,ms2}, in determining the time evolution of $b$ in early eras.  During inflation in RVM ,  the fermion chiral matter fields are assumed \emph{not} to have  been generated; condensation of primordial {\it chiral}~ (i.e. left-right \emph{asymmetric}) gravitational waves (GW) can lead to a condensate of the gravitational CS anomaly term~\cite{Alexander:2004us,Alexander:2006lty,Lyth:2005jf,Mavromatos:2022xdo}:
 \begin{align}\label{condensateN2}
\langle R_{\mu\nu\rho\sigma} \, \widetilde R^{\mu\nu\rho\sigma} \rangle_{\rm condensate\, \mathcal N} 
=\frac{\mathcal N(t)}{\sqrt{-g}}  \, \frac{1.1}{\pi^2} \, 
\Big(\frac{H}{M_{\rm Pl}}\Big)^3 \, \mu^4\, \frac{\dot b(t)}{M_s^{2}} \equiv n_\star \, \frac{1.1}{\pi^2} \, 
\Big(\frac{H}{M_{\rm Pl}}\Big)^3 \, \mu^4\, \frac{\dot b(t)}{M_s^{2}}~,
\end{align} 
where $H$ is the Hubble volume and remains approximately constant during inflation;~in a weak gravity setting graviton modes are integrated over up to an ultraviolet (UV) cut-off $\mu$  to produce the above condensate. In a string-inspired effective field theory framework it is natural to identify $\mu$ with the string scale $M_s$. The dot above the field $b$ denotes a cosmic-time derivative in the Robertson-Walker frame, and $n_\star \equiv \frac{\mathcal N(t)}{\sqrt{-g}} $ denotes the time-dependent number density (over the proper de Sitter volume)  of the sources of GW. We will take $\mathcal N(t)$ to be (approximately) time-independent during inflation. 

The condensate \eqref{condensateN2} will produce a linear axion-$b$ potential in the effective action during inflation. According to the model of \cite{bms1,ms1} only bosonic fields from the massless string gravitational fields in the action 
\eqref{sea6} are present. Such a linear term is reminiscent of models with axion monodromy in conventional string theory~\cite{McAllister:2008hb,Silverstein:2008sg}. The analysis of \cite{bms1} has shown that, as a result of the equations of motion of the $b(x)$ in the bosonic part of \eqref{sea6} we have 
\begin{align}\label{beom}
\partial_\mu b = \sqrt{\frac{2}{3}}\,
\frac{\alpha^\prime}{96\, \kappa} \, {\mathcal K}^\mu,
\end{align}
which, for non-zero condensate value of the temporal component of the topological current $\mathcal K ^0 \simeq {\rm constant} \ne 0$ during inflation, where isotropy and homogeneity holds, implies the solution
\begin{align}\label{bdotconst}
\dot b \equiv \dot{\overline b} = \frac{\alpha^\prime}{96\, \kappa} \, {\mathcal K}^0 
\simeq {\rm constant} \ne 0\,.
\end{align}
We stress that such an axion configuration is linear in (cosmic) time and satisfies the conformal invariance conditions in non-critical strings to be a consistent string background (see \cite{Antoniadis:1988vi}).

As shown in \cite{bms1}, during inflation one can guarantee a consistent solution with $\mathcal K^0 \simeq {\rm constant}$, provided:
 \begin{align}\label{condition2}
1 \simeq  3 \times 10^{-4} \, n_\star \,\Big(\frac{H}{M_{\rm Pl}}\Big)^2 \, \Big(\frac{\mu}{M_s}\Big)^4\quad \Rightarrow \quad 
n_\star^{1/4} \,\frac{\mu}{M_s} \, \sim \,  7.6 \times \Big(\frac{M_{\rm Pl}}{H}\Big)^{1/2}\,.
\end{align} 
The background \eqref{bdotconst} violates Lorentz and CPT symmetry.
 As a consequence, in case the fermions are {\it massive} Majorana RHN, their interaction with the LV and CPTV background will acquire the form \eqref{e2}, where the background axial vector $B_\mu$ of \eqref{temporalB} is given by:
\begin{align}\label{background}
B_\mu = M_{\rm Pl}^{-1} \, \dot{\overline b} \, \delta_{\mu\,0}\,, \quad  \mu=0, \dots 3\,, \quad \dot{\overline b} = {\rm constant}\,,
\end{align}
having only a temporal component. 

If the background $B_\mu$, which remains undiluted until the end of inflation, survives in the epoch of leptogenesis, it would serve as the required axial background. However, in the model of \cite{bms1,bms2,ms1,ms2}, the (chiral) matter fermions appear at the end of the inflationary period as a result of the decay of the unstable inflationary vacuum~\cite{Lima:2013dmf,Perico:2013mna}. This resembles  the  (RVM)~\cite{Shapiro:1999zt}.  In RVM, the primordial CS gravitational anomalies in \eqref{sea6},  may arise during a pre-inflationary phase~\cite{ms1,ms2} as a result of a primordial gravitational wave (GW) background.The source of the GW background  could be collapsing domain walls in theories with dynamically broken supergravities as a consequence of the formation of gravitino condensates. At the end of the inflationary period  the condensates may be {\it cancelled} by the gravitational CS anomalies generated by the chiral fermionic matter at the end of the RVM-like inflationary period. In this case only chiral anomalies containing the gauge fields remain.

That is, one has from \eqref{sea6}:
\begin{align}\label{chiral}
 b(x)  \Big( \sqrt{\frac{2}{3}}\,
\frac{\alpha^\prime}{96\, \kappa} \, {\mathcal K}^\mu_{\,\,\,\,;\mu}
- \frac{\alpha^\prime}{2\, \kappa} \, \sqrt{\frac{3}{2}} \, J^{5\mu}_{\,\,\,\,\,\,\,;\mu} \Big)=  b(x) \, \frac{\alpha^\prime}{2\, \kappa} \, \sqrt{\frac{3}{2}} \, \Big(\frac{e^2}{8\pi^2} \, F_{\mu\nu}(A)\, F^{\mu\nu}(A)  + \frac{g_s^2}{32\, \pi^2} \, \mathcal G_{\mu\nu}^a \, \mathcal G^{a\,\mu\nu} \Big)\,, 
\end{align}
with $e$ the positron charge, $g_s$ the strong interaction coupling, $F_{\mu\nu}(A)$ the Maxwell tensor of 
the photon ($U(1)$) field $A_\mu(x)$, and $\mathcal G_{\mu\nu}^a $, $a=1, \dots 8$, the gluon tensor of QCD. 

In this scenario, chiral gauge anomalies remain in general, as in  \eqref{chiral}. Nonetheless, 
in \cite{bms1,bms2,ms1,ms2} it is assumed that chiral anomalies 
\eqref{chiral} are yet to appear. ( In the scenario of \cite{bms2} chiral anomalies appear at the QCD epoch, see discussion below.) In that case, the equation of motion of the $b$-field  \eqref{beom} has a vanishing right-hand side, since the gravitational anomalies $\mathcal K^\mu$ have been exactly cancelled by the contribution of fermions at the end of inflation. The resulting solution for the KR axion field in that case implies~\cite{bms1} 
\begin{align}\label{T3}
\dot b \propto T^3 \,,
\end{align}
where $T$ is the temperature during the radiation era that succeeds RVM inflation, with the scale factor of the Universe scaling as $a(T) \sim 1/T$.  Because of the relatively short duration of inflation a temperature dependent background of the form \eqref{T3} is considered to be mildly evolving. In such backgrounds,  leptogenesis follows the case of the constant $\dot b$ background of \cite{R9}, a case analysed in detail in \cite{R10Boss,R11Boss}, where it was shown that the basic conclusions of \cite{R9} remain largely valid. Moreover, a full thermal evolution of $B_0$ discussed in  those works has demonstrated that the current value of this LV and CPTV torsion-induced-axion background is well within the pertinent bounds~\cite{Kostelecky:2008ts,Mavromatos:2013gya}

If gravitational and chiral anomalies are absent, for the chiral matter fermions (which are massless above the electroweak phase transition temperature) the axial-fermion-current  interaction term $J^{5\mu}\,\partial_\mu b$ in \eqref{sea6} vanishes as a total derivative . An exception is the axial current of the \emph{massive} RHN, which acquires its mass through various scenarios before the leptogenesis era~\cite{Kostelecky:2008ts,Mohapatra}. A massive RHN is necessary
for our unconventional leptogenesis framework~\cite{deCesare:2014dga,Ellis:2013gca, R10Boss,R11Boss,R12Mav,R13Mav}.\footnote{One may also generate (Majorana) RHN masses dynamically by assuming, as in \cite{R3.13}, the existence of shift-symmetry-breaking couplings of the axions with the RHN, assumed to be generated by non-perturbative effects, say stringy instantons, and as such are suppressed.}   So, even in the absence of anomalies, the RHN equations of motion yield:
 \be\label{axialcurrmassive}
 \partial_{\mu } J^{5\mu}_{\quad ;\mu}=2\,i\,m_{N}\bar{N}\, \gamma^5 \, N\,, \quad m_N \ne 0\,,
 \ee
which is used in the leptogenesis scenario of \cite{R10Boss,R11Boss} to produce an asymmetric decay of RHN 
into standard model lepton and antileptons, in the presence of the background \eqref{T3}.
 \footnote{For completeness we note that, {\it during inflation} in multiaxion models with axions from compatification~\cite{Mavromatos:2022yql},  periodic potentials may be generated for the compactfication axions, as a consequence of world-sheet non-perturbative (instanton) effects at an energy  scale $\Lambda_1 $. Such structures co-exist with the linear-KR-axion potential due to the chiral-GW-induced CS condensates \eqref{condensateN2}. In principle such periodic effective potentials may result in observable modification of the profiles of GW during the radiation era, due to an enhanced production of primordial black holes during the RVM inflationary era~\cite{Mavromatos:2022yql}. It should be also stressed that in some class of these RVM models, depending on the parameters, one may have prolonged reheating periods during the decay of the inflationary vacua, which may contribute further to the above effects.}

\subsection{Generic Einstein-Cartan theory with quantum torsion}\label{EC}

We have seen that~the \emph{totally antisymmetric} KR field 
leads to a totally antisymmetric torsion. In  $(3+1)$-dimensional spacetime this case can be generalised to generic torsion  in the  Einstein-Cartan theory~\cite{Cartan2001RiemannianGI, Kibble:1961ba, Utiyama:1956sy}.
In order to avoid a proliferation of indices it is again convenient to use the language of differential forms. There are two independent 1-forms :
\begin{equation}
\label{E9}
{\bf{e}}^{a}\equiv e^{a}_{\mu }\left( x\right)  dx^{\mu },\  \  \ {\boldsymbol{\omega}}^{a}_{\,\,\, b} \equiv \omega^{a}_{\,\,\, b\mu } \left( x\right)  dx^{\mu }.
\end{equation}
Latin indices denote tangent space indices at a point $x^\mu$, $\mu=0, \dots 3$ of a four-dimensional manifold. There are two associated 2-forms.  The 2-forms ${\bf{T}}^{a}$ for torsion and ${ \bf{R}}^{a}_{b}$ for curvature are given by:

\begin{equation}
\label{E10}
{\bf{T}}^{a}\equiv d{\bf{e}}^{a}+{\boldsymbol{\omega}}^{a}_{\,\,\,b} \wedge {\bf{e}}^{b},\  \  \ { \bf{R}}^{a}_{\,\,\,b}=d{\boldsymbol{\omega}}^{a}_{\,\,\,b} +{\boldsymbol{\omega}}^{a}_{c} \wedge {\boldsymbol{\omega}}^{c}_{\,\,\,b} .
\end{equation}
The \emph{contorsion} 1-form ${\bf{K}}^{ab}$ is by definition 
\begin{equation}
\label{E11}
{\bf{K}}^{ab}\equiv{\boldsymbol{\omega}}^{ab}-\widetilde {\boldsymbol{\omega}}^{ab}
\end{equation}
where $\widetilde {\boldsymbol{\omega}}^{ab}$ is the spin connection in the \emph{absence} of torsion

Since $d{\bf{e}}^{a}+\tilde{{\boldsymbol{\omega}} }^{a}_{\  b} \wedge {\bf{e}}^{b}=0$, $${\bf{T}}^{a}={\bf{K}}^{a}_{b}\wedge {\bf{e}}^{b}.$$
In ECM~\cite{Cartan2001RiemannianGI} the gravitational part of the action on a spacetime manifold $\mathcal {M}$ is written as 
\begin{equation}
\label{E12}
S_{EC}=\frac{1}{4\kappa^{2}}\int \varepsilon_{abcd}\mathbf{\mathcal{R}}^{ab}\wedge\mathbf{e}^{c}\wedge\mathbf{e}^{d}
\end{equation}
The torsion tensor can be decomposed into irreducible representations as follows:
\begin{equation}
\label{E13}
T_{\mu \nu \rho }=\frac{1}{3} \left( T_{{}\nu }g_{\mu \rho }-T_{\rho }g_{\mu \nu }\right)  -\frac{1}{3!} \epsilon_{\mu \nu \rho \sigma } S^{\sigma }+q_{\mu \nu \rho }
\end{equation}
with $\epsilon_{\mu \nu \rho \sigma } q^{\nu \rho \sigma }=q^{\nu }_{\mu \nu }=0$\footnote{We are free to switch between Greek and Latin indices as we choose.}. It is $S^{\sigma }$ which couples to the axial fermion current through the following term in the action~\cite{Duncan:1992vz}:
\begin{equation}
\label{E14}
-\frac{3}{4} \int {\bf{S}}\wedge {\bf{^{*}J}}^{5}\in S_{\psi }\,,
\end{equation}
as follows directly from the fermion kinetic term 
\begin{align}\label{dirac}
\mathcal L_{\rm Dirac} = \bar \psi i \gamma^\mu \mathcal D_\mu (\omega) \, \psi \,, \quad 
\mathcal D_\mu (\omega) = \mathbf 1 \, \partial_\mu  + \frac{i}{8} \, \omega_{\mu\, b}^{a} \, [\gamma_a, \, \gamma^b]\,,
\end{align}
upon using properties of products of three $\gamma^a$ matrices. 
In the string framework this same coupling to the axial current was present cf. \eqref{sea6}.

The contorsion tensor has the following decomposition:
\begin{equation}
\label{E15}
K_{abc}=\frac{1}{2} \epsilon_{abcd} S^{d}+\hat{K}_{abc} .
\end{equation}
$\hat{K}$ includes the trace vector $T_{\mu}$ and the tensor $q_{\mu\nu\rho}$ parts of the torsion tensor.~$S_{EC}$ is written in terms of the Einstein Ricci scalar and contorsion as follows\begin{equation}
\label{E16}
S_{G}=\frac{1}{2\kappa^{2} } \int d^{4}x\sqrt{-g} \left( R+\hat{\Delta } \right)  +\frac{3}{4\kappa^{2} } \int {\bf{S}}\wedge {\bf{S^{\ast}} }
\end{equation}
where $\hat{\Delta } =\hat{K}^{\lambda }_{\  \  \mu \nu } \hat{K}^{\nu \mu }_{\  \  \  \lambda } -\hat{K}^{\mu \nu }_{\  \  \  \nu } \hat{K}^{\  \  \  \lambda }_{\mu \lambda } .$ 

In string theory the KR excitation is expected to have quantum fluctuations, the gravitational axion, since the excitation appears in the spectrum of the quantum string. In our effective field theory for the axion we quantise the torsion (but keep the gravitational field classical) by using path integrals.
 From the classical equations of motion we find
\begin{equation}
\label{E17}
K_{\mu ab}=-\frac{1}{4} e^{c}_{\mu }\epsilon_{abcd} \, \bar{\psi } \gamma_{5} \gamma^{d} \psi 
\end{equation}
which implies 
\begin{align}\label{bianEC}
d^{\ast}{\bf{S}}=0\,,
\end{align} 
and that $Q=\int {\ast}S$ is conserved.\footnote{The reader should notice that \eqref{bianEC} is the analogue of the 
Bianchi identity \eqref{modbianchi2} of the string-inspired case, discussed in the previous subsection \ref{sec:CSmodels}.}
In parallel to the implementation of the Bianchi identity in the string case, we require that this geometric conservation is maintained at the \emph{quantum} level by introducing a factor $\delta \left( d^{\ast }S\right) $ in the path integral (using a  canonically normalised pseudoscalar  Lagrange multiplier field $b(x)$).\footnote{ Maintaining this constraint in the quantum theory will require appropriate counterterms~\cite{Duncan:1992vz}.} The \emph{torsion part} of the quantum gravity path integral is
\begin{eqnarray}\label{E18}
Z & = & \int DSDb\exp \left[ i\int \frac{3}{4\kappa^{2} } S\wedge \ast S-\frac{3}{4} S\wedge \ast J^{5}+\left( \frac{3}{2\kappa^{2} } \right)^{1/2}  b~d\ast S\right] \nonumber \\
 & = & \int Db\exp \left[- i\int \frac{1}{2} db\wedge \ast db+\frac{1}{f^{\rm EC}_{b}} db\wedge \ast J^{5}+\frac{1}{2 (f^{\rm EC}_{b})^2} J^{5}\wedge \ast J^{5}\right]  
\end{eqnarray} 
where we have performed the Gaussian integration over  $S$ in the second line. Consequently the Lagrangian multiplier field becomes the dynamical axion field. By contrast, in the string case, the target-space effective action, which involves the torsion field $\mathcal H_{\mu\nu\rho}$,  
 is an \emph{infinite} series in powers of the Regge slope $\alpha^\prime $~\cite{Green:2012oqa,Gross:1986mw,Metsaev:1987zx}. Only on truncating this series up to and including  $\mathcal O(\alpha^\prime)$ terms, the  resulting effective action is quadratic in the torsion field $H_{\mu\nu\rho}$, which results in \eqref{sea6}. The higher order in $\alpha^\prime$ terms involve higher powers and higher derivatives of the $\mathcal H$ field, and so cannot be integrated analytically. This aspect makes the Einstein-Cartan framework valuable and simpler for the study of KR axions in leptogenesis.

 The torsion-induced axion coupling in the ECM is
\begin{align}\label{eccoupl}
f^{\rm EC}_{b}=\sqrt{\frac{8}{3}} \frac{1}{\kappa} \approx 4\times 10^{18}~{\rm GeV}\,. 
\end{align}
 We remark that the fermionic part of \eqref{E18} coincides with that of \eqref{sea6}, for  $\alpha^\prime = \kappa^2$. The totally antisymmetric part of the torsion $\varepsilon_{\mu\nu\rho\sigma}\, S^\sigma$ corresponds to the $H_{\mu\nu\rho}$-torsion in the string theory model. However, the string-inspired theory \eqref{sea6} of \cite{bms1,ms1} is also characterised by the presence of \emph{primordial gravitational anomaly terms} which may be cancelled by chiral matter generated after inflation\cite{bms1,ms1}. 
  
In the generic Einstein-Cartan theory the gravitational and gauge anomaly terms arise only at the one-loop level, due to  the axial current (see  \eqref{anom}, below). 
Indeed, we may partially integrate the second term in the exponential in \eqref{E18} and note that the axial current has anomalies, both gravitational or gauge~\cite{Duncan:1992vz, Adler:1969gk, Fujikawa:1979ay, R12Mav}:
\begin{align}\label{anom} 
d{\bf{^{\ast }J^{5}}}=-\frac{\alpha_{QED} Q^{2}}{\pi } {\bf{F}}\wedge {\bf{F}}-\frac{\alpha_{s} N_{q}}{2\pi } Tr\left( {\bf{G}}\wedge {\bf{G}}\right)  -\frac{N_{f}}{8\pi^{2} } {\bf{R}}^{ab}\wedge {\bf{R}}_{ab}
\end{align}
where $\alpha_{QED}$ and $\alpha_{s}$ are the fine structure constants for (electromagnetic) Abelian  and (quantum chromodynamic) non-Abelian gauge fields, respectively; $N_f (N_q)$ is the number of fermion (quark only) flavours and $Q^{2}=\sum_{f} Q^{2}_{f}$ where $Q_f$ is the electric charge of the fermion with flavour $f$; ${\bf{F}}$, ${\bf{G}}$ and ${\bf{R}}$ are the field strength 2-forms of QED, QCD and gravity (with torsion).\footnote{In the string-inspired case of 
\eqref{sea6}, it can be shown that the anomaly does not contain torsion parts, as the latter can be removed by appropriate counterterms~\cite{Hull:1985dx, Mavromatos:1987ru} (see also \cite{R13Mav}).} 

\subsection{Comparative phenomenology of the string inspired  and Einstein-Cartan models}
In the string-inspired models (RVM) of \cite{bms1,ms2,ms1}, in the very early Universe, an era of primordial gravitational-wave dominance is speculated and may lead to the gravitational anomaly term to condense:
\begin{align}\label{condEC}
\langle\frac{N_{f}}{8\pi^{2} } {\bf{R}}^{ab}\wedge {\bf{R}}_{ab}\rangle \simeq {\rm constant}\,,
\end{align} 
via the mechanism of \cite{Alexander:2004us, Alexander:2006lty, Lyth:2005jf}. 

During inflation, the Hubble parameter is approximately constant $H \simeq H_I ={\rm constant}$, and, therefore, in the presence of this condensate, the equation of motion of the axion $b$ field stemming from the Einstein-Cartan action \eqref{E18} implies:
\begin{align}\label{beomEC}
\Box\, b \simeq  \frac{1}{f_b^{\rm EC}} \, \langle\frac{N_{f}}{8\pi^{2} } {\bf{R}}^{ab}\wedge {\bf{R}}_{ab}\rangle \simeq {\rm constant}\,.
\end{align}
If we assume homogeneous and isotropic fields, with dependence only on cosmic time $t$, $b = b(t)$ 
during inflation; we may approximate the above equation as (the overdot denotes derivative with respect to $t$) 
\begin{align}\label{beomECcosm}
\ddot b(t) + 3 H_I \, \dot b \simeq {\rm constant}\,,
\end{align}
which gives the solution (for large times)
\begin{align}\label{bdotback}
\dot b \simeq \frac{1}{3H_I \, f_b^{\rm EC}}\, \langle\frac{N_{f}}{8\pi^{2} } {\bf{R}}^{ab}\wedge {\bf{R}}_{ab}\rangle\, \simeq {\rm constant}\,.
\end{align}
This background \eqref{bdotback} remains undiluted until the end of inflation and into the early radiation era.

After leptogenesis in the Einstein-Cartan model, there is most likely no
 primordial gravitational wave dominance and no condensate \eqref{condEC}. On account of \eqref{beomECcosm}, there is 
an ordinary cosmic evolution for the (massless) field $b$ in the radiation era.

We next remark that the case \eqref{E18}, besides leptogenesis, has other possible uses: providing (i)  axionic dark matter~\cite{Preskill:1982cy,Abbott:1982af,Dine:1982ah,Adams:2022pbo,Marsh:2015xka,Mehta:2021pwf} and (ii) a solution of the strong CP problem~\cite{Peccei:1977hh}. The resulting effective action  is:
\begin{eqnarray}
\label{E21}
S_{eff}&=&S_{0}-\frac{1}{2(f^{\rm EC}_{b})^2} \int J^{5}\wedge \ast J^{5}-\frac{\alpha_{QED} }{\pi f^{\rm EC}_{b}} \int bF\wedge F\\
& &-\int \frac{1}{2} db\wedge \ast db-\frac{1}{8\pi^{2} } \int \left( \Theta +\frac{N_{f}}{f^{\rm EC}_{b}} b\right)  R^{ac}\wedge R_{ac}\nonumber\\
&& -\frac{\alpha_{s} }{2\pi } \int \left( \theta+\frac{N_{q}}{f^{\rm EC}_{b}} b\right)  Tr\left[ G\wedge G\right] \nonumber 
\end{eqnarray}
The constants $\Theta$ and $\theta$ are associated with topological terms and have been kept in for completeness (and classically do not affect the equations of motion).~$S_{0}$ represents the Einstein gravitational interactions, non-gravitational  gauge interactions and Yukawa interactions of matter. This model embodies the links with axion cosmology, and the $b$ field is at the heart of the connections.
Using the action in \eqref{e2a} and \eqref{e2} we have explored leptogenesis. Recently the formulation used in \eqref{E21} has also been investigated in connection with  axionic dark matter. Although the model has features similar to that of the QCD axion, the gravitational axion couples to the Pontryagin densities~\cite{Nakahara:206619} of \emph{all} the gauge interactions in the model. The model could, for example, be a generalisation of the SM above the electroweak transition; then all the gauge fields would have couplings to the axion through their Pontryagin density. Another feature is the repulsive current -current interaction. This would be a contribution to the de Sitter nature of the Universe which is however suppressed by the factor of $(f^{\rm EC}_b)^2$ (\eqref{eccoupl}). There have been attempts to link condensates of such repulsive axial current-current terms to the cosmological constant~\cite{Poplawski:2010jv}.

\subsubsection{The axion model}
Having linked, under certain conditions, the Einstein-Cartan model to the CPTV and LV leptogenesis scenario of \cite{R9,R10Boss,R11Boss}, with axion backgrounds $\bar b$ satisfying $\dot \bar{b} \simeq {\rm const.}$, 
we can  return to the full path integral and quantum fluctuations of the axion field. We write 
\begin{equation}
\label{ee18}
Z=\int \mathcal{D}g\mathcal{D}\psi \mathcal{D}\bar{\psi } \mathcal{D}b\exp \left[ iS_{eff}\right]  .
\end{equation}
In realistic situations with many fermion species $i$, we have:
\begin{equation}
\label{Ee1}
{}_{}J^{5}_{\mu }=\sum^{f}_{i=1} \bar{\psi }_{i} \gamma_{\mu } \gamma^{5} \psi_{i} .
\end{equation}
We will split the $b$ field into a quantum  $\widetilde b$ and classical part $\bar b$; the effective action (in conventional notation) can be written as 
\begin{eqnarray}
\label{e2a}
S_{eff} & = & \frac{1}{2\kappa^{2} } \int d^{4}x\, \sqrt{-g} \, \left( R+\frac{8}{3} \partial_{\sigma } \, \bar{b} \,\partial^{\sigma }  \bar{b} -\Omega \right)  +S_{free}-\int d^{4}x\, \sqrt{-g} \, \partial_{\mu } \bar{b} \, J^{5\mu } \nonumber\\
&  & -\frac{3\kappa^{2} }{16} \int d^{4}x \, \sqrt{-g} \, J^{5}_{\mu }\, J^{5\mu }+\frac{8}{3\kappa^{2} } \int d^{4}x\, \sqrt{-g} \, \partial_{\sigma } \bar{b} \  \partial^{{}^{\sigma }} \widetilde{b} 
 \nonumber\\
 &  & +\frac{1}{2\kappa^{2} } \int d^{4}x\,\sqrt{-g} \, \frac{8}{3} \partial_{\sigma } \widetilde{b} \, \partial^{\sigma } \widetilde{b} +\frac{1}{\kappa }\int d^{4}x\,\sqrt{-g} \,\widetilde{b}\,  \partial_{{}\mu } J^{5\mu }
\end{eqnarray}
 where $S_{free}$ is the action  of a free fermion in a gravitational background, where we added a potential cosmological constant, $\Omega$, for the sake of generality. This background decouples from $\widetilde b$ since the $\bar b \widetilde b$ term vanishes (provided $\int d^{4}x\  \partial^{0} \bar{b}$ vanishes).\footnote{The constant background could have a small effect in the effective theory at temperatures lower than the mass of the sterile neutrino since it affects the propagator of the sterile neutrino.} The term $\int d^{4}x\,\sqrt{-g}\, \partial_{\mu } \bar{b}\, J^{5\mu }$ gives the model of leptogenesis represented by \eqref{e2} with a constant axial-vector background \eqref{temporalB}.

\section{Effective axion potential and metastability 
 in the CPTV leptogenesis model }\label{sec:stable}

In our UV complete theory (see ~\eqref{e2a}) the axion has a kinetic term and a Yukawa-like interaction only . The sterile neutrino has a heavy mass $m_N$ and so at energies much lower than $m_N$, we can determine the effective potential for the KR axion\cite{Coleman:1973jx,Manohar:2020nzp,Burgess:2020tbq, Georgi:1993mps}. In order for our model  to have implications for dark matter~\cite{Profumo:2017hqp} and the strong CP problem~\cite{Dine:1986bg}, it is necessary to determine the stability of the axion vacuum from the effective potential (using nonperturbative semiclassical analysis on this potential) for our unconventional leptogenesis models \cite{deCesare:2014dga,R10Boss,R11Boss}. 

Although the effective field theory can be obtained through analysis of Feynman diagrams~\cite{R7.4},  it is more convenient to consider a path integral formulation where the heavy particle fields can be integrated out in the post inflationary action \eqref{sea6}. This results in a local effective potential. When the heavy particle is a fermion, this procedure has been studied for a large class of theories relevant for BSM physics and leads to the fermionic universal effective action and is tabulated~\cite{Ellis:2020ivx,Zhang:2016pja}. 
  This procedure for heavy fermion particles  leads from a UV complete Lagrangian $\mathcal{L}$ to an effective Lagrangian  $\mathcal{L_{\rm{eff}}}$,  which is schematically given by
\begin{equation}
\label{ee1}
\int DND\bar{N} \exp \left( i\int d^{D}x\  {\mathcal{L}}\left( \tilde{b} ,N,\bar{N} \right)  \right)  =\exp \left( i\int d^{D}x\  {\mathcal{L_{\rm{eff}}}}\left( \tilde{b} \right)  \right).    
\end{equation}

In \emph{both} Einstein-Cartan and RVM theories, the coupling of the torsion-induced axion to the divergence of the axial current, has the  form
\begin{align}\label{rhnaxial}
\mathcal S^{\rm {b-J}^5}_{\rm int} = - \frac{1}{f_b} \, \int d^x \, \sqrt{-g} \, \widetilde{b}(x) \,  J^{5\mu}_{\quad ;\mu}\,,
\end{align}
where $f_b$ is the axion coupling, given by \eqref{kraxcoupl}(\eqref{eccoupl}) for RVM (Einstein-Cartan) theories. 

In \eqref{e2a} (and also \eqref{E21}) the heavy fermion is the sterile RHN fermion; we explicitly consider the contribution to \eqref{rhnaxial} from the sterile neutrino.
The presence of the RHN mass 
implies a {\it shift-symmetry breaking}; hence we expect that an integration over the RHN field will induce an effective potential for the gravitational axion, which could imply a {\it metastability} of the axion vacuum in the Einstein-Cartan (and  RVM) frameworks in the post-inflationary era\footnote{There are independnet mechanisms for the mass of the gravitational axion, which have been investigated using Schwinger-Dyson analyses~\cite{R3a}  and modifications of  arguments for the mass of the QCD axion in a different era of the Universe~\cite{Holdom:1982ex,Castillo-Felisola:2015ema}.}.

In the notation of~\cite{R12a} the interaction of RHN has the strucuture
\be
\label{interaction}
\bar{N} \  \left( \slashed p-m_{N}-X[\tilde{b} ]\right)  N,
\ee
where $X$ is parametrised in~\cite{Ellis:2020ivx}~as 
\be
X\left[ \tilde{b} \right]  =W_{0}\left[ \tilde{b} \right]  +iW_{1}\left[ \tilde{b} \right]  \gamma^{5} +V_{\mu }\left[ \tilde{b} \right]  \gamma^{\mu } +A_{\mu }\left[ \tilde{b} \right]  \gamma^{\mu } \gamma^{5}. 
\ee
In our case the relevant part of $X\left[ \tilde{b} \right]$ is 
\be
W_{1}\left[ \widetilde{b} \right]  =\frac{2m_{N}}{f_{b}}  \widetilde{b}.
\ee
The resultant effective axion potential (up to dimension $6$ operators~\footnote{\label{f17}Using~\cite{Coleman:1973jx} we note that
 the subdominant dimension 8 operator gives a positive contribution to the effective potential.}) is:
\be\label{effpotb}
V_{\rm eff} [b] = a_{2}\left( W_{1}\left[ \widetilde{b} \right]  \right)^{2}  +a_{4}\left( W_{1}\left[ \widetilde{b} \right]  \right)^{4}  +a_{6}\left( W_{1}\left[ \widetilde{b} \right]  \right)^{6}  
\ee
where 
\be\label{axmass}
a_{2}=4m^{2}_{N}\left( 1-\frac{1}{2} \ln \frac{m^{2}_{N}}{\mu^{2} } \right) \,,
\ee
  \be
  a_{4}=\frac{5}{6} -\ln \frac{m^{2}_{N}}{\mu^{2} } \,,
  \ee
  and
  \be\label{instab}
  a_{6}=-\frac{1}{3m^{2}_{N}}\, .
  \ee
  In the above formulae, $\mu$ is a transmutation mass scale 
  in the effective theory defined by the heavy RHN mass scale $m_N \sim 10^5$~GeV in the model of \cite{deCesare:2014dga,R10Boss,R11Boss}. We note that in our framework one has to have  $f_{b} \gtrsim m_N$, for our effective field theory to be valid.
  
  In our case the large mass $m_N$  plays the r\^ole of an ultraviolet cutoff for the effective theory.  One may do a matching~\cite{Kaplan:2005es}  at\footnote{We note, for completeness, that the $\mu$-dependent logarithms should be considered as in~\cite{Zhang:2016pja} to include poles in the dimensional regularisation with spacetime dimensions  $d = 4 -\epsilon$, $\epsilon \to 0+$ , i.e. the poles can  be recovered by the replacement
   \begin{align}\label{muepsilon} 
 -{\rm ln}\Big(\frac{m_N^2}{\mu^2}\Big) \rightarrow \frac{1}{\epsilon}  -{\rm ln}\Big(\frac{m_N^2}{\mu^2}\Big)\,,
 \end{align} Hence in the $\overline{\rm MS}$  regularisation scheme, where only the residues of the poles $\frac{1}{\epsilon}$ are kept in the analysis, one can implement the matching \eqref{mN=mu} safely. A more detailed dependence of the effective potential on $\mu$ may be studied with the renormalisation group~\cite{Einhorn:2007rv}, when we embed our toy theory in a more phenomenological realistic effective low-energy theory obtained from the underlying string theory model. This will not be the topic of discussion in this work.}  
 \begin{align}\label{mN=mu}  
  m_N \simeq \mu\,,
  \end{align}
 and run the scale $\mu$ in the regime 
  \begin{align}\label{mNmu} 
  \mu \lesssim m_N \,,
  \end{align}
to lower values until one hits an experimentally measurable physical mass scale.~In our system, such a mass scale would be the axion $b$ mass $m_b$. The latter is generated by the integration of the heavy RHN, which induces an effective KR axion mass given by: 
 \begin{align}\label{axionmass}
  m_b^2 = 32\, \frac{m_N^4}{f_b^2}\, \Big(1 - \frac{1}{2}{\rm ln}[\frac{m_N^2}{\mu^2}]\Big)\,.
  \end{align}
  \begin{itemize}
      \item For the string model \eqref{sea6}, with the maximum allowed value of $ f_b \sim 2 \times 10^2 \, \kappa^{-1} $( for the case $\alpha^\prime \sim \kappa^2$ ({\it cf.} \eqref{kraxcoupl}), and $\mu \sim m_N \ll f_b$ of order $10^5$~GeV (as in the model of \cite{R9,R10Boss,R11Boss}), we obtain from \eqref{axionmass} axion masses of order 
$m_b \sim 0.14 $~eV. This mass is \emph{too large} for the KR axion in this model to play the role of the QCD axion (of $\mathcal O(10)~\rm{\mu eV}$), in contrast to the expectations of \cite{bms2, Castillo-Felisola:2015ema}.
 
  \item However, 
this mass is within the range of an axion-like-particle (ALP) dark matter component~\cite{Marsh:2015xka,Mehta:2021pwf}. Indeed, an acceptable ALP mass range, consistent with astrophysical and cosmological constraints~\cite{Mehta:2020kwu}, is: $10^{-20} \, {\rm eV} \lesssim m_{\rm ALP} \lesssim {\rm eV}$. Hence, due to \eqref{kraxcoupl}, the upper bound on the ALP mass is quite restrictive for the string scale in such models, allowing only models with large string mass scales close to ({\it i.e.} at most a couple of orders away from) the Planck mass. This is also consistent with the Einstein-Cartan model \eqref{eccoupl}.
\item
We observe from \eqref{axionmass}, that, as the scale $\mu$ runs towards smaller values within the regime \eqref{mNmu}, 
 $m_b^2$ is non-negative provided 
  \begin{align}\label{posaxmass}
  {\rm ln}\Big(\frac{m_N}{\mu}\Big) \le 1 \quad \Rightarrow \quad \mu \gtrsim e^{-1}\, m_N\,,
  \end{align} 
 where the saturation of the inequality implies zero axion mass generation by this mechanism.
Therefore we cannot run $\mu$  to values lower than $e^{-1}\, m_N\,,$ if we do not want to have tachyonic behaviour~\cite{etde_7242567} in the axion sector.\footnote{Of course here we are assuming that there are no large additional contributions to the mass of the axion from other mechanisms which could off-set any negative contribution. In particular, we note that because of the coupling of the axion to the gauge Chern-Simons anomalous interactions  as mentioned above, there are contributions to the axion effective potential from instantons in Quantum Chromodynamics,. These contributions, however, lead to a very small positive axion mass, thus not affecting our conclusions.} As follows from \eqref{axmass}, this result is independent of the magnitude of the field $b$. Taking into account \eqref{mNmu}, we thus obtain the allowed regime of running of  $\mu$:
\begin{align}\label{muallowed}
e^{-1} m_N \, \lesssim \, \mu \, \lesssim \, m_N\,.
\end{align}

 \item 
 In the region \eqref{posaxmass} (and hence also for \eqref{mN=mu}), the effective potential \eqref{effpotb} appears stable up to order $b^4$ in the axion self-interaction.  
 From \eqref{instab}, however, we observe that the potential is unstable at sixth order in powers of the field $b(x)$, for sufficiently large fields $b^2 > f^2_b$. Following methods applied to the study of the (meta)stability of a (false) vacuum~\cite{Coleman:1977py, R34cc}, such as the Higgs boson under the assumption of new physics~\cite{Lee:1985uv,Isidori:2001bm,Branchina:2013jra}, we may calculate the tunnelling time of an unstable vacuum by first looking for the bounce solution (tree level in the effective theory)~\cite{Coleman:1977py} to the Euclidean equation of motion, and then computing the quantum fluctuations on top of the classical solution~\cite{R34cc}.
  \end{itemize}
 \begin{figure}[t]
 \centering
  \includegraphics[clip,width=0.60\textwidth,height=0.30\textheight]{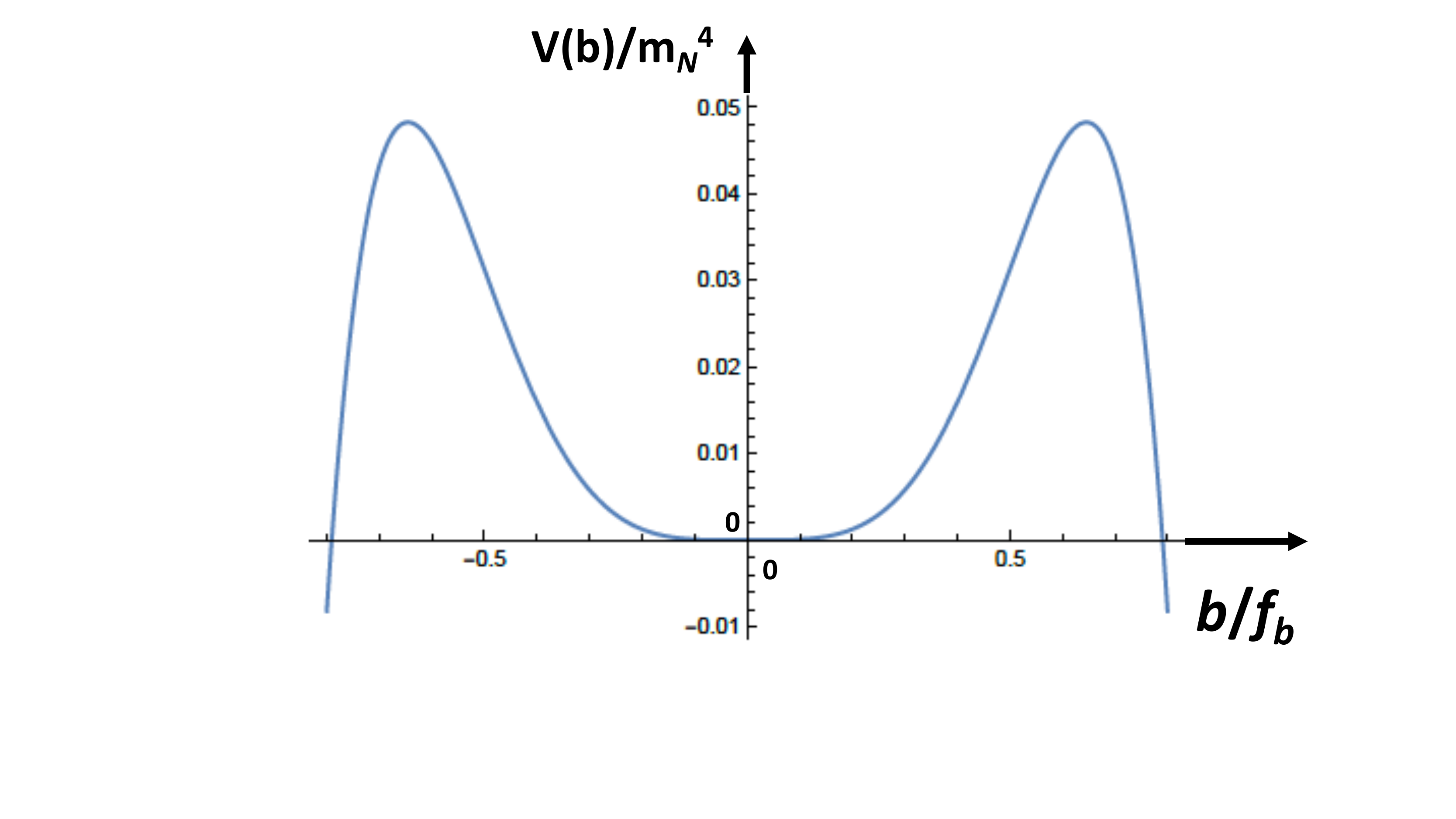} 
\caption{\it The quartic and sixth-order interactions in the effective axion potential \eqref{effpotblambda} in units of $m_N$ vs. $b/f_b$ at $\mu=m_N$ ($V[b]/(2m_N)^4= \, (b/f_b)^4 \, (5/6 - (4/3)\, (b/f_b)^2)$. The second derivative of the potential vanishes at the local minimum $<b>=0$, which leads to the flatenning of the potential around that minimum. The corresponding Euclidean potential, obtained upon applying \eqref{reverse}, corresponds to a Higg-like potential, where the local maxima $b_{\rm max} = \pm \eta$ in the figure correspond to non-trivial minima, at which the bounce solutions are calculated. }\label{fig:pot}
\end{figure} 
 From the Higgs-potential analogue in~\cite{Branchina:2013jra},\footnote{The potential \eqref{effpotb} has even powers of the field, hence the pseudoscalar nature of the field is not relevant for our considerations here. The effective potential here resembles the effective potential appearing in the analysis of the (meta)stability of the Higgs vacuum; we follow that analysis in determining the stability of the axion potential.} we may rewrite the effective potential 
 \eqref{effpotb} for relatively large values of the field $b$, relevant for the metastability argument, as: 
 \begin{align}\label{effpotblambda}
 V_{\rm eff}[b] \simeq \frac{\lambda_{\rm eff}(b, \mu)}{4}\, b^4 \,, \quad \lambda_{\rm eff}(b, \mu) \equiv  \frac{2^6\, m_N^4}{f_b^4}\, \Big[ \frac{5}{6}- {\rm ln}(\frac{m_N^2}{\mu^2})  - \frac{4}{3\, f_b^2}\, b^2 \Big]\,. 
 \end{align}
 
 Under assumption \eqref{mNmu}, for values of $b$ such that 
 \begin{align}\label{bgrtfb}
 \frac{b^2}{f_b^2} > \frac{5}{8} - \frac{3}{2}{\rm ln}\Big(\frac{m_N}{\mu}\Big)\,,
 \end{align}
the effective self-interaction coupling $\lambda_{\rm eff}$ defined in \eqref{effpotblambda} changes sign from positive to negative, thus triggering a potential instability (see fig.~\ref{fig:pot}). In the $\overline{\rm MS}$ scheme, in which \eqref{mN=mu} is valid, 
the instability arises for values of $b \gtrsim 0.79\, f_b$.

To estimate the false vacuum lifetime, we remark  that we write the effective potential in the form \eqref{effpotblambda}, and reverse the sign of the coupling~\cite{Coleman:1977py,R34cc} 
\begin{align}\label{reverse}
\lambda_{\rm eff} \to  \lambda^{\rm (E)}_{\rm eff} = - \lambda_{\rm eff}
\end{align}
for a $\phi^4 $ Euclidean scalar theory; there are known bounce solutions, used in the estimation of the tunnelling time of the false vacuum~\cite{Coleman:1977py,R34cc}. 
Because of \eqref{reverse} and \eqref{mN=mu} (in the $\overline{{\rm MS}}$ scheme), , the potential \eqref{effpotblambda} of fig.~\ref{fig:pot} is reversed in sign, thus acquiring a Higgs-like form, where the local unstable maxima at  $b= \pm \eta$ ({\it cf.} Fig. \ref{fig:pot}) correspond (in the Euclidean case) to non-trivial minima:
\begin{align}\label{localmax}
\frac{\eta}{f_b} \simeq 0.65\,.
\end{align}
Then, following \cite{Coleman:1977py,R34cc}, the (inverse of the) tunnelling time $\tau$ (lifetime) of the false vacuum is computed as:
\begin{align}\label{tunnellingrate}
\tau^{-1}  = T_U^{-3}\, \frac{1}{4\pi^2} \, S^2[b_b] \Big| \frac{{\rm det}^\prime (-\Box + V^{\prime\prime} (b_b)}{{\rm det}(-\Box + V^{\prime\prime}(<b>))}\Big|^{-1/2}\, e^{-S(b_b)}   \,,
\end{align}
where $<b>$ denotes the vacuum expectation value that minimises the quartic axion potential, in terms of which the bounce is found: in this case $<b>=0$; $b_b= b(r^2)$ (where $r^2=x^\mu_E x^\mu_E$) is the Euclidean (E) solution for the bounce; $S[b_b]$ is the action 
for the bounce; the notation ${\rm det}^\prime$ means that the zero-modes are excluded in the computation of the determinant; a a prime over the potential $V$ denotes derivative with respect to the axion field $b$; the quantity $T_U$ denotes the the age of the Universe and is  estimated by the cosmological data~\cite{Planck} to be 13.8 billion years, or in reduced Planck-mass units
\begin{align}\label{Age}
T_U \simeq 1.6 \times 10^{60} \, M_{\rm Pl}
\end{align}
 The boundary between the metastability and instability regions of the axion potential is $\tau=T_U$, with $\tau > (<) T_U$ indicating a metastable (unstable) vacuum. 

The action of the bounce $S[b_b]$ for the model involving the $b^6$ unstable terms in the potential \eqref{effpotblambda}, 
 can be calculated numerically only~\cite{Branchina:2013jra, Branchina:2014usa}, extending standard methods~\cite{Coleman:1977py,Lee:1985uv}, to obtain:
\begin{align}\label{newphysicsbounce}
S[b^{(1)}_b] \simeq \Big( 1 - (\gamma + 1)^4\Big)\, \frac{8\pi^2}{3|\lambda_{\rm eff}(b_b, \mu)|} \,, 
\end{align}
where 
\begin{align}\label{gammadef} 
\gamma \equiv -\frac{\lambda^{\rm (E)}_{\rm eff}}{\lambda^{(E)} } \, \frac{1}{1 + \frac{\lambda_6^{(\rm E)}}{\lambda^{(\rm E)}} \, \frac{\eta^2}{f_b^2}}\,,
\end{align}
and $\lambda^{(E)} <0$ denotes the coupling of the inverted  standard $b^4$ scalar potential, $V(b)=\frac{1}{4}\lambda^{(\rm E)} \, b^4$, without the $b^6$ corrections, and the coefficient $\lambda_6$ defines the $b^6$ corrections in the (Euclideanised) potential \eqref{effpotblambda}, which (on assuming the $\overline{\rm MS}$-scheme condition \eqref{mN=mu}) we write in the form :
\begin{align}\label{potb6}
V^{(\rm E)}= \frac{5}{6}\,\lambda_0 \, b^4 - \frac{1}{3} \,\lambda_0 \,\frac{b^6}{f_b^2}\,, \quad \lambda_0 \equiv -2^4\,\frac{m_N^4}{f_b^4}\,.
\end{align}
As becomes clear from its form ({\it cf.} the inverted analogue of fig.~\ref{fig:pot}), the Euclidean potential \eqref{potb6} can be well approximated with just the inverted $\phi^4$ potential for $b \lesssim \eta$. Following the general method of \cite{Lee:1985uv}, which was used in  \cite{Branchina:2013jra} to estimate the lifetime of the Higgs (false) vacuum from new physics corrections, it suffices to evaluate the lifetime $\tau$ of the false vacuum associated with the potential \eqref{potb6}, and $\lambda_{eff}$ at the point where the bounce $b_b \sim \eta$, with $\eta$ given by \eqref{localmax}.

From \eqref{potb6}, \eqref{gammadef}, we therefore have that at such values of $b_b$
\begin{align}\label{gammadef2}
\gamma = -\frac{\lambda^{(\rm E)}_{\rm eff}}{\lambda^{(\rm E)}} \, \frac{1}{1 + \frac{\lambda_6^{(\rm E)}}{\lambda^{(\rm E)}}\, \frac{\eta^2}{f_b^2}} \simeq \frac{-1+\frac{8}{5} (\frac{\eta}{f_b})^2}{1-\frac{2}{5} (\frac{\eta}{f_b})^2} \,  \stackrel{\eqref{localmax}}{\simeq} \,  - \frac{0.324}{0.831} \simeq -0.39\,.
\end{align}
According to the (numerical) analysis for the existence of bounce solutions performed in \cite{Branchina:2013jra,Branchina:2014usa}, such solutions do exist, provided $0 > \gamma > -1$, which is satisfied by \eqref{gammadef2} for our axion system.
Indeed, the (approximate) bounce solutions of the Euclidean equations of motion for the $b^6$-corrected potential are given by~\cite{Branchina:2013jra,Branchina:2014usa}:
\begin{align}\label{bounce1}
b^{(1)}_b (r) = \Big\{ \begin{array}{c} 2\eta - \eta^2 \sqrt{\frac{|\lambda_{\rm eff}|}{8}} \, \frac{r^2 + \overline R^2}{\overline R} \quad 0 < r < \overline r\, \\
\sqrt{\frac{8}{|\lambda_{\rm eff}|}}\, \frac{\overline R}{r^2 + \overline R^2} \,\,\,\,\,\,\, \qquad \qquad r > \overline r\,,
\end{array}
\end{align}
where $\overline r^2 = \frac{8\,\gamma}{\lambda^{(\rm E)}_{\rm eff} \eta^2}(1 + \gamma) > 0$, and 
$\overline R^2 = \frac{8}{|\lambda_{\rm eff}|}\, \frac{\gamma^2}{\eta^2}$, which confirms explicitly our earlier assertion that the bounce solution exists only for $0 > \gamma > -1$~\cite{Branchina:2013jra,Branchina:2014usa}. 

The analysis of \cite{Branchina:2013jra}  also shows that there are other approximate bounce solutions for the range of $b \ll f_b$,  which are those of the original $b^4$ scalar potential $V(b)=\frac{\lambda}{4}\, b^4$, of the form:
\begin{align}\label{new}
b^{(2)}_b(r) = \sqrt{\frac{2}{|\lambda_{\rm eff}|}}\, \frac{2R}{r^2 + R^2}\,,
\end{align}
where the size $R$ of these bounces can take any value in the range $\eta^{-1}\, \sqrt{\frac{8}{|\lambda_{\rm eff}|}} < R < \infty$. The Euclidean action of such bounces is independent of $R$ and of the form of the standard scalar $\frac{\lambda}{4}b^4$ potential~\cite{Coleman:1977py}:\footnote{The degeneracy with respect to the size of the bounce is lifted by considering one-loop corrections, which are calculated on top of the classical solutions, and which select one 
particular value for the size.}
\begin{align}\label{standard}
S[b_b^{(2)}] \simeq \frac{8\pi^2}{3|\lambda|}\,.
\end{align} 
The explicit computation of the determinant factors in \eqref{tunnellingrate} for the bounce solution yields the formula for the total width $\Gamma$ of the false vacuum corresponding to \eqref{potb6} (in the concrete case of, say, two (approximate) bounce solutions $b_b$, $b_b^{(2)}$, assumed to have been selected by quantum corrections to the action):
\begin{align}\label{widthf}
\Gamma = \frac{1}{\tau} \sim \sum_{i=1,2}\, \frac{1}{T_U} \, \frac{T_U^4}{R^{(i)4}_b}\, e^{-S[b^{(i)}_b]} \, e^{-\Delta S_i}\,
\end{align}
where $i=1,2$ labels the bounce solutions \eqref{bounce1} and \eqref{new}, respectively, with sizes $R_b^{(i)}$, bounce actions
$S[b_b^{(i)}]$ in  \eqref{newphysicsbounce}, \eqref{standard}, and $\Delta S_i$ the loop corrections to the bounce actions. 
 When the axion potential \eqref{potb6} (or, equivalently \eqref{effpotblambda}) is generated at one loop, we set 
\begin{align}\label{deltas}
\Delta S_i =0\,, \quad i=1.2\,,
\end{align}
to avoid overcounting of the quantum corrections. 

On taking into account \eqref{gammadef2}, 
the actions \eqref{standard} and 
\eqref{newphysicsbounce} of the two bounce solutions are of the same order: 
\begin{align}\label{S1S2}
\frac{S[b_b^{(1)}]}{S[b^{(2)}_b]}  = \Big( 1 - (\gamma + 1)^4\Big)\,\frac{|\lambda|}{|\lambda_{\rm eff}|} =
\frac{ 1 - (\gamma + 1)^4}{1-\frac{2}{5} (\frac{\eta}{f_b})^2} \,
\stackrel{\eqref{localmax}}{\simeq}\, \frac{0.86}{0.83} \, \simeq 1.04\,.
\end{align} 
This differs from the standard Higgs-potential case~\cite{R7.6,Branchina:2013jra,Branchina:2014usa}, where 
the ``new-physics'' bounce solutions (from  powers of the Higgs scalar higher than four), lead to actions that are smaller by several orders of magnitude~\cite{Branchina:2013jra} from the action of the standard $\phi^4$ bounce}. Thus, in our case, such solutions  will not affect the order of magnitude of the estimate  of the lifetime of the false vacuum based only on the bounce solution $b_b$ \eqref{bounce1}. 
Hence we restrict ourselves only to the bounce $i=1$, corresponding to \eqref{potb6}, in \eqref{widthf}, and set the renormalisation group (RG) scale
equal to the size of the bounce,  
\begin{align}\label{mu=R}
\mu=1/R^{(1)}\,.
\end{align}
We select  $\mu$ to maximise the tunnelling probability~\cite{R7.6}.
 Although we have taken the bounce to be the classical Euclidean solution (to the equations of motion corresponding to the effective potential \eqref{potb6} assuming\eqref{mN=mu}), when calculating the lifetime we take into account one-loop corrections, by using running couplings. Explicitly the couplings in the respective formulae are replaced by the running ones evaluated at the point  $b \sim \eta$, with $\eta$ given by \eqref{localmax}, while we set the RG scale equal to the bounce size \eqref{mu=R} that minimizes the lifetime. 
 
On account of \eqref{deltas}, we consider only the bounce $i=1$ in \eqref{widthf}, and set $\Delta S_1=0 $. Hence, we evaluate the (minimal) lifetime $\tau$ at a bounce size which maximises the probability of tunnelling of the false vacuum, using the formula: 
\begin{align}\label{tunnel} 
 \tau = T_U \, \min_{\mu} \, \mathcal T(\mu)\,, \quad \mathcal T(\mu) \sim T_U^{-4}\, \mu^{-4}
 \, \exp\Big(\frac{8\pi^2}{3|\lambda_{\rm eff}(\eta, \mu)|}\Big)\,.
 \end{align} 
 Here $\mu$ is the running RG scale at the value that minimises the quantity $ \mathcal T(\mu)$~\cite{Branchina:2013jra,Branchina:2014usa}, which is 
 indicated by the ``$\displaystyle{\min_{\mu}}$'' symbol in \eqref{tunnel}.  
  
 The running of the $\lambda_{\rm eff} $ with the RG scale requires further analysis if one wishes to include higher-loop corrections to the masses and wave-function renormalization of the KR axion field $b(x)$; our analysis suffices since we are interested only in the leading estimate of the tunnelling time, compared with $T_U$. 
 
 We observe that, in the case that the effective coupling $\lambda_{\rm eff}$ in \eqref{effpotblambda} is {\it positive} ($\lambda_{\rm eff} > 0$), it 
increases with the RG scale $\mu$, since: 
 \begin{align}\label{lamrg} 
\frac{d\lambda_{\rm eff}}{d{\rm ln}\mu} = 2^7 \, \frac{m_N^4}{f_b^4} > 0\,, 
\end{align} 
on assuming bare quantities for $m_N$ and $b$, i.e. ignoring mass and wave-function renormalization effects to leading order \footnote{The latter depend on the underlying detailed microscopic model for the KR axion  which we ignore in this work.}.

Let us suppose, for concreteness, $f_b \gg m_N$ ({\it e.g.} either generic Einstein-Cartan torsion models \eqref{eccoupl}, or string-inspired models in which $\alpha^\prime \sim \kappa^2$). In such cases, at an initial UV cut-off scale $\mu_0$ of the effective theory, which can be set to be the heavy RHN mass $\mu_0 = m_N\sim 10^5$~GeV of the CPTV leptogenesis model of \cite{R9}, the effective coupling $\lambda_{\rm eff}$ (in the Minkowski-spacetime effective potential) takes on the value:
\begin{align}\label{leff}
\lambda_{\rm eff} (b \simeq \eta, \mu=m_N) \simeq  17.3  \times \Big(\frac{m_N}{f_b}\Big)^4 \ll 1 \,, 
\end{align}
with the scale $\eta$ being given by \eqref{localmax}. Thus, due to \eqref{lamrg}, $\lambda_{\rm eff} (\mu)$  becomes smaller as we decrease $\mu$, until $\mu$ reaches a value for which $\lambda_{\rm eff}$ {\it vanishes}, and then becomes {\it negative}.

From \eqref{tunnel} we then observe \emph{formally} that when $\lambda_{\rm eff}    > 0$ there is {\it no minimum} of $\tau$, unless $\mu \to \infty$, in which case, as we observe from \eqref{effpotblambda} and \eqref{tunnel}, $\tau \to 0$ (we assume that the  fields $b < f_b$, are such that $\lambda_{\rm  eff}$ does not vanish). Indeed, for $\lambda_{\rm eff} >0$, we have 
\begin{align}\label{derl}
\frac{d}{d\mu}\Big(\frac{1}{\lambda_{\rm eff}}\Big) = - \frac{1}{\lambda^2_{\rm eff}} \frac{1}{\mu} \frac{2^7\, m_N^4}{f_b^4} < 0\,.
\end{align}
Thus, passing onto the Euclidean spacetime to compute quantities associated with the lifetime $\tau$, we have (with the index $(E)$ denoting Euclidean quantities):
\begin{align}\label{dtaumu}
\frac{\delta \tau}{\delta \mu} = - 4\, T_U^{-4}\, \mu^{-5} \, e^{\frac{8\pi^2}{3|\lambda_{\rm eff}^{(E)}|}} + T_U^{-4}\, \mu^{-4} \,\frac{8\pi^2}{3}\, \frac{d}{d\mu}\Big(\frac{1}{|\lambda_{\rm eff}^{(E)}|}\Big)\, e^{\frac{8\pi^2}{3|\lambda_{\rm eff}^{(E)}|}} < \, 0\,, \quad {\rm on~account~of~}\eqref{derl}\,.
\end{align}
On the other hand, 
\begin{align}\label{d2taumu}
\frac{\delta^2 \tau}{\delta \mu^2} &= 20 \, T_U^{-4}\, \mu^{-6} \, e^{\frac{8\pi^2}{3|\lambda_{\rm eff}^{(E)}|}} 
- 4\,  T_U^{-4}\, \mu^{-5} \,\frac{8\pi^2}{3}\, \frac{d}{d\mu}\Big(\frac{1}{|\lambda_{\rm eff}^{(E)}|}\Big) \, e^{\frac{8\pi^2}{3|\lambda_{\rm eff}^{(E)}|}} + T_U^{-4}\, \mu^{-5} \,\Big[\frac{8\pi^2}{3}\, \frac{d}{d\mu}\Big(\frac{1}{|\lambda_{\rm eff}^{(E)}|}\Big) \Big]^2 \, e^{\frac{8\pi^2}{3|\lambda_{\rm eff}^{(E)}|}} \nonumber \\
&- 4\,  T_U^{-4}\, \mu^{-5} \,e^{\frac{8\pi^2}{3|\lambda_{\rm eff}^{(E)}|}}\, \frac{8\pi^2}{3}\, \frac{d}{d\mu}\Big(\frac{1}{|\lambda_{\rm eff}^{(E)}|}\Big) \, > \, 0\,, \quad {\rm on~account~of~}\eqref{derl}\,.
\end{align}
Thus, there is a \emph{monotonic decrease} of $\tau(\mu)$ in this case, with the (trivial) minimum occurring for $\mu \to \infty$, for which $\tau =0$.

Minimisation of $\tau$, on the other hand, occurs for negative $\lambda_{\rm eff} < 0$.
Indeed it can be readily seen that the mimimum of $\mathcal T (\mu)$ occurs for 
$\mu=\mu_{\rm min}$ such that:
\begin{align}\label{mumin}
1 & = \frac{2^8\, \pi^2}{3\, \lambda^2_{\rm eff}(b, \mu_{\rm min})} \frac{m_N^4}{f_b^4} \quad 
\Rightarrow \quad \lambda^{(-)}_{\rm eff} (b, \mu_{\rm min}) = - \frac{2^4\, \pi}{\sqrt{3}} \frac{m_N^2}{f_b^2} \quad
\Rightarrow \quad {\rm ln}\Big(\frac{m_N^2}{\mu^2_{\rm min}}\Big) = \frac{5}{6} + \frac{\pi}{4\sqrt{3}}\, \frac{f_b^2}{m_N^2}
- \frac{4}{3}\frac{b^2_b}{f_b^2}\,.
\end{align}

To estimate the $\mu_{\rm min}$ in \eqref{mumin} we need to know an estimate of the bounce solution of the effective potential \eqref{effpotblambda}. Following \cite{Branchina:2014usa,Lee:1985uv} we set $b_b \sim \eta$ (\eqref{localmax}) in the estimation of $\lambda_{\rm eff}$ that enters the formula \eqref{tunnel} for the computation of the false-vacuum life time. In the full theory, to which the axion system under consideration is embedded, one needs to compute 
loop corrections to the running $\lambda_{\rm eff}(\eta,\mu)$, and use such complete expressions in the estimate of the life time $\tau$ using \eqref{tunnel}. 
 
From \eqref{mumin} we observe that for the CPTV leptogenesis model  of \cite{R9}, for which $f_b \sim \mathcal O(M_{\rm Pl}) \gg m_N \sim 10^5$~GeV
the minimum $\mu_{\rm min} \ll m_N$, approaching zero for all practical purposes (infrared fixed point). Such minima correspond to very-large-size $\overline R \sim \mu_{\rm min}^{-1}$ instantons. In such a regime of $\mu$, the lifetime of the vacuum \eqref{tunnel} can be very large (practically infinite), as follows from  the very small value ({\it cf.} \eqref{mumin}) of $|\lambda^{(-)}_{\rm eff} (b, \mu_{\rm min})| \ll 1$, for $m_N \ll f_b$ which characterises the model of leptogenesis of \cite{R9}:
\begin{align}\label{wrongvac}
\frac{\tau(\mu = \mu_{\rm min})}{T_U} \sim 6.5 \times 10^{-240}\, \Big(\frac{M_{\rm Pl}}{m_N}\Big)^4 \, e^{\frac{5\pi}{4\sqrt{3}} \, \frac{f_b^2}{m_N^2}} \,.
\end{align}
In 
this region of $\mu$ the KR axion acquires a negative mass square, which, in combination with the fact that the entire axion potential \eqref{effpotb} is negative in this case, leads to a stable situation in which the KR axion appears as a {\it tachyon} matter field with 
negative unstable potential energy. Such a phase is \emph{unphysical}, and thus we discard it. Nonetheless,  it is worth  noting that in our simplified system the tachyon appears as a metastable phase of the leptogenesis model of \cite{R9}, with a lifetime much longer than the age of the Universe, $ \tau(\mu = \mu_{\rm min}) \gg T_U$, provided $f_b \gg m_N$, which is the case of either Einstein-Cartan torsion theories \eqref{eccoupl}, or large-scale string theories with $\alpha^\prime \sim \kappa^2$ \eqref{kraxcoupl}. However, for low string mass scales $\sqrt{\alpha^\prime} = M_s^{-1} \gg \kappa = M_{\rm Pl}^{-1}$, for instance, such that $f_b \simeq m_N$ \eqref{kraxcoupl}, we see that the vacuum is highly unstable, with $\tau \ll T_U$.

On the other hand, for $\mu=m_N$, which is the regime of interest to us, if we want to generate dynamically an axion mass in a regime relevant to DM interpretations of the KR axion,  the lifetime of the corresponding false vacuum can be computed from \eqref{tunnel} to yield:
\begin{align}\label{lifetimemu=mN}
\frac{\tau}{T_U} &\sim (m_N T_U)^{-4} \, \exp(\frac{8\pi^2}{3|\lambda^{(E)}_{\rm eff}(\eta, \mu=m_N)|}) \, \stackrel{\eqref{leff}}{\simeq} \, 6.5  \times 10^{-240} \, \Big(\frac{M_{\rm Pl}}{m_N}\Big)^4 \, \exp\Big[\frac{8\pi^2}{51.9} \Big(\frac{f_b}{m_N}\Big)^4\Big]  \nonumber \\
&\simeq 1.6 \times 10^{-185} \, \exp\Big[1.5 \Big(\frac{f_b}{m_N}\Big)^4\Big]\,,
\end{align}
where we took into account the age of the Universe \eqref{Age}.\footnote{On the other hand, for very large (transplanckian) field values 
$b \gtrsim f_b \gg m_N$ such that  
\begin{align}\label{transpl}
\frac{5}{6} + \frac{\pi}{4\sqrt{3}}\, \frac{f_b^2}{m_N^2}
- \frac{4}{3}\frac{b^2_b}{f_b^2}\, < 0,
\end{align}
in the right-hand-side of the last equality of \eqref{mumin}, we observe that
the minimum of $\mathcal T (\mu)$ occurs within the desired regime \eqref{mNmu}, in which the one-loop-induced effective mass term of the axion $b$ field is a proper mass, and the vacuum is again metastable, given that $\tau \gg T_U$ (we remark that we are agnostic though as to whether such transplanckian field values make sense in an UV complete theory of quantum gravity, such as string theory~\cite{Ooguri:2006in,Palti:2019pca,Garg:2018reu,Ooguri:2018wrx,Denef:2018etk,Martin:2000xs}.)}
For the cases of generic Einstein-Cartan torsion theories \eqref{eccoupl}, or large-scale string theories  ($\alpha^\prime \sim \kappa^2$ \eqref{kraxcoupl}), this yields clearly a metastable vacuum, with $\tau \gg T_U$.~From \eqref{lifetimemu=mN},~for $\sqrt{\alpha^\prime} \gg \kappa$, for instance, such that $f_b \simeq m_N$ \eqref{kraxcoupl}, we see that the vacuum is highly unstable, with $\tau \ll T_U$.

It would be interesting to consider the effect of dimension-eight ($\propto b^8$), or even higher-order, operators in our analysis, and establish 
whether there is a second local minimum in the corresponding effective potential. (In the example of new physics for the Higgs potential examined in \cite{Branchina:2013jra,Branchina:2014usa}, and discussed above, this was the case.) We could then find a minimum $\mu_{\rm min} \sim m_N$. Whether such a situation is realised, remains to be seen although we expect this to be the case in view of the analysis in \cite{Coleman:1973jx,Manohar:2020nzp} at one loop for the effective potential. Indeed, that work implies that 
 the subdominant dimension 8 operator gives a positive contribution due to logarithmic corrections to the quartic scalar self-interaction (see footnote \ref{f17}).

From the above discussion, we conclude that for the regime of parameters of the models we study, one faces a metastable vacuum for the KR  axion $b$, with $<b>=0$, 
in string-inspired models with $f_b \gg m_N$. 
In view of its acquired mass \eqref{axionmass}, of $\mathcal O(0.14~{\rm eV})$ for concrete string models \eqref{sea6} with $\alpha^\prime \sim \kappa^2$, or generic Einstein-Cartan theories \eqref{eccoupl}, 
the KR axion could play the r\^ole of an axion-like-particle (ALP) DM component~\cite{Preskill:1982cy,Abbott:1982af,Dine:1982ah,Marsh:2015xka}, in a range consistent with astrophysical and cosmological constraints~\cite{Mehta:2020kwu}. We remark, though, that, due to \eqref{kraxcoupl}, the upper bound on the ALP mass is quite restrictive for the string scale in such models, allowing only models with large string mass scales close to ({\it i.e.} at most a couple of orders away from) the Planck mass. This is also consistent with the generic Einstein-Cartan torsion models, \eqref{eccoupl}).
  
\section{Conclusions}\label{sec:concl}
  
 In this work we have established formally a connection between contorted geometry and particle dark matter by exploiting the role of torsion as axion-like particles in a model for CPTV leptogenesis with heavy RHN. Specifically, we have considered two frameworks in which torsion induced axions develop an effective potential as a result of integrating out these heavy sterile fermions.
 \begin{enumerate}
     \item The first framework is the Einstein-Cartan theory in which the axion couples to massless and massive fermions. The chiral anomalies associated with chiral (massless) fermions imply in such theories a coupling of the geometric axion with gauge and gravitational  Chern-Simons terms with a dimensionful coupling constant of the order of the Planck scale.

     \item The second framework is a string-inspired gravitational theory, in which the role of torsion is played by the field strength of an antisymmetric tensor spin one field of the massless gravitational multiplet of the string. This behaves as a totally antisymmetric component of the torsion, which in $3+1$ dimensions is dual to an axion like particle, which shares similar couplings to Chern-Simons anomalies to the Einstein-Cartan framework. This framework contains many  degrees of freedom that contribute to the phenomenology in addition to the torsion induced axion compared to the first framework.
 \end{enumerate}
  
 Using the methods of effective field theory, we find a one-loop quantum potential for our axion at energies low compared to the RHN mass $m_N$. The potential is negative for large enough values of the field and so the issue of vacuum stability of the axion field arises. We have found that the stability of the axion vacuum depends on the ratio of the axion coupling $f_b$ to $M_N$, ${f_{b}}/{m_{N}} $ as seen in \eqref{lifetimemu=mN}. It should be stressed that our effective field theory is valid for ${f_{b}}/{m_{N}} \gtrsim 1$. 
 We found that for high values of $f_b/m_N \gg 1$, the vacuum is metastable with a life time much longer than the age of the Universe. In such cases, the axions acquire an effective mass which makes them credible DM candidates.
 On the other hand, a ratio $f_b/m_N \sim 1$ is not viable for our model, as it leads to very short vacuum life times. 

In our approach we have considered for simplicity truncation of the one-loop effective axion potential to dimension six operators. It would be desirable to explore the effective potential to all powers of the axion field, using the methods of Coleman and Weinberg~\cite{Coleman:1973jx}. We shall also investigate gradient effects in the calculation of the tunnelling rate to the true vacuum~\cite{PhysRevD.92.125022}, due to the spatial inhomogeneities of the bounce solution.

\section*{Acknowledgments}

We acknowledge discussions with Sebastian Ellis, Ken Mimasu and Tevong You. 
The work of N.E.M. and S.S. is supported in part by the UK Science and Technology Facilities research
Council (STFC) and UK Engineering and Physical Sciences Research
Council (EPSRC) under the research grants ST/T000759/1 and  EP/V002821/1, respectively. N.E.M.  also acknowledges participation in the COST Association Action CA18108 ``{\it Quantum Gravity Phenomenology in the Multimessenger Approach (QG-MM)}''. We thank Carl Bender for a useful discussion.

\appendix

\section{Lifetime estimate using the full potential}
 
 We supplement the numerical calculations in Sec.\ref{sec:stable} with related analysis in which we include the mass terms when finding the maximum of the potential terms (cf. \eqref{localmax}). The method is semi-analytic but this additional refinement does not change any conclusion. We work in terms of the variable $\epsilon$ where $\frac{m^{2}_{N}}{\mu^{2} } =1+\epsilon $ and $\epsilon   <0$.
 In terms of  $\hat{b} \equiv \frac{b}{f_{b}} $ and  $\frac{m^{2}_{N}}{\mu^{2} } =1+\epsilon $ (where $\left| \epsilon \right|  <1$)  the rescaled effective potential   Eq\eqref{effpotblambda} is 
 \begin{equation}
\label{es1}
\tilde{V}_{eff} \left( \hat{b} ,\epsilon \right) \equiv \frac{V_{eff}\left( \hat{b} \right)  }{m^{4}_{N}} 
= \left( 1-\frac{1}{2} \log \left( 1+\epsilon \right)  \right)  \hat{b}^{2} +\left( \frac{5}{6} -\log \left( 1+\epsilon \right)  \right)  \hat{b}^{4} -\frac{4}{3} \hat{b}^{6} 
\end{equation}
 and the effective quartic coupling in the Euclidean formulation is
 \begin{equation}
\label{es2}
\lambda_{eff} \left( \hat{b} ,\epsilon \right)  \equiv  -\frac{2^6\, m_N^4}{f_b^4}\,\Big[ \left( \frac{5}{6} -\log \left( 1+\epsilon \right)  \right)  -\frac{4}{3} \hat{b}^{2}\Big]\,. 
\end{equation}
 The characteristic  size of the potential  is estimated from the condition
 \begin{equation}
\label{es3}
 \frac{d\tilde{V}_{eff} }{d\hat{b} } =0
\end{equation} 
 which will also be relevant in the estimation of the lifetime of the KR axion vacuum. In terms of $u=\hat{b}^{2} $ the solution of \eqref{es3}  is $u=u_{\pm }(\epsilon)$ where 
 \begin{equation}
\label{es4}
u_{\pm }(\epsilon)=\frac{1}{24} \left( 5-6\log \left( 1+\epsilon \right)  \pm \sqrt{g\left( \epsilon \right)  } \right)  
\end{equation}
and 
\begin{equation}
\label{es5}
g\left( \epsilon \right)  =169-132\log \left( 1+\epsilon \right)  +36\  \left( \log \left( 1+\epsilon \right)  \right)^{2}.  
\end{equation}
 The maximum of $V_{eff}$ occurs for  $\hat{b}(\epsilon)=\eta(\epsilon)=\sqrt{u_{+}\left( \epsilon \right)  } $ (cf. \eqref{localmax}).   Moreover,  Pad$\acute{\rm{e}}$  approximants represent $\lambda_{eff} \left( \eta,\epsilon \right)$  well
 \begin{equation}
\label{es6}
\lambda_{eff} \left( \eta,\epsilon \right)\approx -  \frac{2^6\, m_N^4}{f_b^4}\,h(\epsilon)\end{equation}
where $$h(\epsilon)=\frac{0.0003 \epsilon
   ^6+0.0027 \epsilon ^5-0.1076
   \epsilon ^4-0.6219 \epsilon
   ^3-1.1399 \epsilon ^2-0.7990
   \epsilon -0.1667}{0.0003
   \epsilon ^6-0.0071 \epsilon
   ^5+0.0418 \epsilon ^4+0.6836
   \epsilon ^3+2.1239 \epsilon
   ^2+2.4866 \epsilon +1}\,.$$
for small $\epsilon$ \footnote{The use of Pad$\acute{\rm{e}}$  approximants leads to poles for $\frac{1}{\lambda_{eff}}$ of which all but one are spurious and not in the region of physical interest and so the approximation is useful~\cite{R15}. 
.} .The lifetime $\tau(\epsilon)$ of the axion vacuum is approximated by 
 \begin{equation}
\label{es7}
\frac{\tau \left( \epsilon \right)  }{T_{U}} \approx A^{-1}\left( 1+\epsilon \right)^{2}  \exp \left( \frac{\pi^{2} }{24} \left( \frac{f_{b}}{m_{N}} \right)^{4}  |h\left( \epsilon \right)|^{-1}  \right)  
\end{equation}
where $A=\left( T_{U}m_{N}\right)^{4}  $. On minimising $\tau \left( \epsilon \right) $ we find $\epsilon \approx -.8$ which requires $\mu \approx 2.24\  m_{N}$. We recall that the mass scale $\mu$ is used in dimensional regularisation to make the couplings dimensionless. Let us reexamine the scaling properties of the tunnelling (bounce) configuration (for which we do not have an analytic form). The bounce is localised in a spacetime volume of size $L^4$ where $L\simeq \frac{2\pi }{k} $ and will be denoted by $b_k$. $L$ is small for a bounce and hence there is a contribution from large $k$. The field configuration $b_k$  consists of a wave packet with large $k$. The action $S_E$ for this configuration is
\be
S_{E}\approx \left( 2\pi \right)^{4}  \left[ \frac{\hat{b}^{2}_{k} }{2} +\frac{m^{2}_{b}}{2} \hat{b}^{2}_{k} +\frac{\lambda }{4!} \hat{b}^{4}_{k} +\sum^{\infty }_{n=1} \left( c_{n}\left( \frac{k^{2}}{m^{2}_{N}} \right)^{n}  \hat{b}^{4+2n}_{k} +\ldots \right)  \right].  
\ee
By choosing $\mu^{2} \sim k^{2}\sim O\left( m^{2}_{N}\right) $ we would probe strongly the $b^6$ term ($n=1$) which is responsible for tunnelling and this is consistent with the minimising criterion for the lifetime.

\bibliography{bibli2.bib}

\end{document}